\documentclass[onecollarge,natbib]{svjour2}
\bibpunct{[}{]}{;}{n}{}{,} 
\smartqed  
\usepackage{graphicx}
\usepackage{mathptmx}      
\usepackage{amsmath}
\usepackage{amssymb}

\usepackage{color}
\newcommand{\non}{\nonumber\\}

\def\quad{\hskip1.5em\relax}

\newcommand{\sll}{\raise.15ex\hbox{$/$}\kern-.43em\hbox{$l$}}
\newcommand{\slepsilon}{\raise.15ex\hbox{$/$}\kern-.53em\hbox{$\epsilon$}}
\newcommand{\slvarepsilon}{\raise.15ex\hbox{$/$}\kern-.53em\hbox{$\varepsilon$}}
\newcommand{\slL}{\raise.15ex\hbox{$/$}\kern-.53em\hbox{$L$}}
\newcommand{\slP}{\raise.15ex\hbox{$/$}\kern-.53em\hbox{$P$}}
\newcommand{\slp}{\raise.1ex\hbox{$/$}\kern-.63em\hbox{$p$}}
\newcommand{\slq}{\raise.1ex\hbox{$/$}\kern-.53em\hbox{$q$}}
\newcommand{\slv}{\raise.1ex\hbox{$/$}\kern-.63em\hbox{$v$}}
\newcommand{\slR}{\raise.15ex\hbox{$/$}\kern-.53em\hbox{$R$}}
\newcommand{\slQ}{\raise.15ex\hbox{$/$}\kern-.53em\hbox{$Q$}}
\newcommand{\slK}{\raise.15ex\hbox{$/$}\kern-.53em\hbox{$K$}}
\newcommand{\slk}{\raise.15ex\hbox{$/$}\kern-.53em\hbox{$k$}}
\newcommand{\slSigma}{\raise.15ex\hbox{$/$}\kern-.53em\hbox{$\Sigma$}}
\newcommand{\slcalP}{\raise.15ex\hbox{$/$}\kern-.63em\hbox{$\cal P$}}
\newcommand{\slA}{\raise.15ex\hbox{$/$}\kern-.73em\hbox{$A$}}
\newcommand{\slbfA}{\raise.15ex\hbox{$/$}\kern-.73em\hbox{${\imb A}$}}
\newcommand{\slpartial}{\raise.15ex\hbox{$/$}\kern-.53em\hbox{$\partial$}}
\newcommand{\sla}{\raise.15ex\hbox{$/$}\kern-.53em\hbox{$a$}}
\newcommand{\slb}{\raise.15ex\hbox{$/$}\kern-.53em\hbox{$b$}}
\newcommand{\slc}{\raise.15ex\hbox{$/$}\kern-.53em\hbox{$c$}}
\newcommand{\slD}{\raise.15ex\hbox{$/$}\kern-.53em\hbox{$D$}}
\newcommand{\slC}{\raise.15ex\hbox{$/$}\kern-.53em\hbox{$C$}}

\journalname{Few Body Systems}
\begin{document}

\title{Quarkonium production at collider energies in Small-$x$ formalism
\thanks{Invited presentation at the workshop "New Observables In Quarkonium Production", ECT$^\star$, Italy, 28 February - 4 March 2016.}
}


\author{Kazuhiro Watanabe}


\institute{K. Watanabe \at
              Key Laboratory of Quark and Lepton Physics (MOE) and Institute of Particle Physics, Central China Normal University, Wuhan 430079, China \\
              \emph{Present address: Physics Department, Old Dominion University, Norfolk, VA 23529, and Theory Center, Jefferson Lab, Newport News, VA 23606}\\  
              \email{watanabe@jlab.org}
}

\date{Received: date / Accepted: date}

\maketitle

\begin{abstract}
I present a short review of recent studies of quarkonium production in proton-proton and proton-nucleus collisions at collider energies in Small-$x$ formalism.

\end{abstract}

\section{Introduction}\label{sec:1}

\quad
Heavy quark pair production in proton-proton (pp) and proton-nucleus (p$A$) collisions at high scattering energies has been studied to investigate nonlinear gluon saturation dynamics~\cite{Gribov:1984tu,Mueller:1985wy,Mueller:2001fv} inside hadron and nucleus. Gluon saturation is an universal phenomenon which can appear when nonlinear gluon recombination effect is no longer negligible at extremely small value of the Bjorken $x$ in hadron and nucleus. In the deep saturation regime of hadron and nucleus, a transverse size of the gluon is characterized by the saturation scale $Q_s$ inversely. Empirically, the saturation scale for nucleus can be estimated by $Q_{sA}^2\sim A^{1/3}x^{-0.3}$ with $A$ the atomic mass number. As an important consequence, the saturation scale for heavy nuclei in the RHIC and LHC energies is comparable with heavy quark mass. Therefore, the study of heavy quark pair production in p$A$ collisions has been an intriguing issue in the small-$x$ physics community.

At present, abundant data about heavy quark pair production are available in pp and p$A$ collisions at RHIC and LHC. Among many observables, quarkonium production has been studied actively, since the transverse momentum distribution of quarkonium production must reflect the saturation dynamics. Thanks to the effective factorization between heavy quark pair production and bound state formation, we can compare quantitatively theoretical calculations with the current available data. 

Heavy quark pair production in pp/p$A$ collisions has been addressed in the Color-Glass-Condensate (CGC) framework or Small-$x$ saturation formalism~\cite{Weigert:2005us,Gelis:2010nm,Kovchegov:2012mbw}. In the small-$x$ formalism, p$A$ collision and pp collision in the forward rapidity region can be regarded as a dilute-dense system in which small-$x$ (large-$x$) gluons for target nucleus (projectile proton) participate in hard partonic scatterings. Therefore, rapidity dependence of transverse momentum gluon distributions for proton and nucleus is important and can be calculated numerically by means of JIMWLK equation~\cite{Jalilian-Marian:1997jx,JalilianMarian:1997gr,Iancu:2001ad,Iancu:2000hn,Weigert:2000gi} or Balitsky-Kovchegov (BK) equation~\cite{Balitsky:1995ub,Kovchegov:1996ty}.

In the meantime, as we mention in this paper, description of bound state formation at long distance depends on model. Although elemental quarkonium production mechanism is still a challenge of QCD~\cite{Brambilla:2004wf,Brambilla:2010cs}, the next three kinds of models have been incorporated in the small-$x$ formalism for phenomenology; Color Evaporation Model~\cite{Fujii:2006ab,Fujii:2013gxa,Ducloue:2015gfa}, Color Singlet Model~\cite{Kharzeev:2005zr,Dominguez:2011cy}, and Color Octet Model or Non-Relativistic QCD factorization approach~\cite{Kang:2013hta,Ma:2014mri,Ma:2015sia}. An interesting point is that nuclear dependence of differential cross section for quarkonium production in p$A$ collisions in the small-$x$ formalism can be universal when large-$N_c$ approximation is assumed. Thus, the small-$x$ formalism has the power to predict nuclear modification factor of quarkonium production in p$A$ collisions.

Study of the nuclear modification of quarkonium production in p$A$ collisions is usually expressed as exploration of cold nuclear matter (CNM) effects, such as nuclear shadowing~\cite{Eskola:2009uj,Hirai:2001np,Vogt:2015uba}, nuclear absorption~\cite{Ferreiro:2008wc,Ferreiro:2013pua}, energy loss in cold nucleus~\cite{Arleo:2012hn,Arleo:2012rs,Arleo:2013zua}, comover interaction~\cite{Gavin:1996yd,Ferreiro:2014bia}, and the gluon saturation. Since quarkonium production has been considered as a good probe to examine hot QCD medium created in high energy heavy ion collisions~\cite{Andronic:2015wma}, precise examination of the CNM effects is the essential subject in heavy ion physics. 

Thus far, many theoretical studies have developed individual scenario of the CNM effects and combined a couple of the effects together. However, a combination of the saturation effect with the other effects has been hardly yet considered. Recently, incorporating the saturation effect with the energy loss mechanism has begun but not completed~\cite{Munier:2016oih}. Although we will restrict our attention to the gluon saturation effect on quarkonium production in this paper, we must consider all of the CNM effects together to interpret data quantitatively.

This paper is aimed at reviewing the recent studies of quarkonium production from the small-$x$ formalism point of view and clarifying problems which must be resolved in the future. This paper is organized as follows. In Section~\ref{sec:2}, we consider how we should factorize effectively between partonic scattering part at short distance and bound state formation part at long distance from the small-$x$ formalism viewpoint. Subsequently, we review the small-$x$ formalism for describing heavy quark pair production in p$A$ collision in Section \ref{sec:3} and \ref{sec:4}. In Section~\ref{sec:5}, we revisit basic concepts of quarkonium production models. In Section~\ref{sec:6} and \ref{sec:7}, we compare the selected results computed in the small-$x$ formalism with data at RHIC and LHC.


\section{Probing the gluon saturation}\label{sec:2}

\quad
In this section, we consider a concept of factorization between short distance part and long distance part for quarkonium production in p$A$ collisions in the small-$x$ formalism. The factorization allows us to use safely quarkonium production in p$A$ collisions as a valuable probe into the gluon saturation inside high energy nucleus. 

We first follow the discussion about time scales separation in target rest frame~\cite{Kharzeev:2005zr}. In the rest frame of target nucleus, interaction time of proton scattering off nucleus is characterized by $\tau_{int}\sim {R_A}$ in natural unit. $R_A$ is size of target nucleus. Heavy quark pair ($q\bar q$) is produced in p$A$ collisions over the time scale $\tau_P\sim \frac{1}{2m_q}\frac{E_g}{2m_q}$ where $m_q$ is quark mass and $E_g$ is energy of incident gluon which subsequently splits into the $q\bar q$. Due to momentum conservation, one finds $(2m_q)^2=x_px_As=2x_px_AM_NE_p$ with $x_{p}$ ($x_{A}$) being longitudinal momentum fraction of projectile proton (target nucleus) carried by incident gluon. $M_N$ is mass scale of nucleon and $E_g=x_pE_p$. Therefore, one can find $\tau_P\sim \frac{1}{2x_A M_N}$. 

At high scattering energy or forward rapidity (proton going direction), $\tau_P$ is much larger than $\tau_{int}$ owing to Lorentz time dilation. Indeed, $x_{p,A}$ can be determined from observables in final state by using $x_{p,A}=e^{\pm y}\sqrt{M^2+P_\perp^2}/\sqrt{s}$. For example, $J/\psi$ production at low transverse momentum provides $10^{-2}\lesssim x_A \lesssim10^{-3}$ at RHIC in the forward rapidity and $10^{-4}\lesssim x_A \lesssim 10^{-5}$ at LHC in the forward rapidity. The large scale of $\tau_{int}$ suggests that projectile proton interacts coherently with target nucleus. In other words, the $q\bar q$ pair is produced coherently in high energy p$A$ collision. The coherent interaction between the $q\bar q$ and many gluons in the target nucleus can reflect important information on the gluon saturation dynamics. 

Regarding time scale of quarkonium formation (e.g. $J/\psi$), we can estimate it as $\tau_F\sim \frac{2}{M_{\psi(2S)}-M_{J/\psi}}\frac{E_g}{M_{J/\psi}}$ with $M_{J/\psi}$ being mass of $J/\psi$ and $M_{\psi(2S)}$ being mass of $\psi(2S)$. Since binding energy of quarkonium must be much smaller than $m_q$, one can find immediately $\tau_F\gg \tau_P$. Thus, for quarkonium production in high energy p$A$ collisions, we obtain
\begin{align}
\tau_F\gg \tau_P\gg \tau_{int}.
\label{eq:scale-separation-rest-frame}
\end{align}
Eq.~(\ref{eq:scale-separation-rest-frame}) expresses that dynamics of bound state formation can be decoupled from target nuclear matter effect. However, as discussed in Ref.~\cite{Qiu:2013qka}, the factorization Eq.~(\ref{eq:scale-separation-rest-frame}) could be a subtle in the center of mass frame in p$A$ collision. Therefore, let us elaborate the above discussion again in the center of mass frame in p$A$ collision. 

As mentioned in Section~\ref{sec:1}, transverse momentum ($P_\perp$) distribution of quarkonium production in p$A$ collisions should reflect the saturation dynamics. Therefore, when we restrict ourself to quarkonium production at very low $P_\perp$, we can assume $\Lambda_{\text{QCD}}\ll P_\perp\sim Q_{sA}\ll M$ in the partonic hard scattering level. Here $M$ is quarkonium mass and $Q_{sA}$ is the saturation scale of the gluon inside target nucleus. The $q\bar q$ pair gains $P_\perp\sim Q_{sA}$ from the multiple scattering in target nucleus.

Next we consider a characteristic scale of the bound state formation process. Thus far, Color Evaporation Model (CEM) and Non-Relativistic QCD (NRQCD) approach have succeeded in describing data of quarkonium production in pp collisions at collider energies. Therefore, now we suppose that we can apply these approaches to quarkonium production in p$A$ collisions. In NRQCD and CEM, relative velocity of heavy quark ($v$) and strong coupling constant are the essential expansion parameters for long distance matrix elements~\cite{Bodwin:2005hm}. Therefore, the momentum of the produced quark ($m_qv$) can characterize the bound state formation. From the factorization viewpoint, long distance dynamics which is expressed by $m_qv$
must be decoupled from short distance dynamics which is embedded in $Q_{sA}$. However, if $Q_{sA}\sim m_qv\sim Mv/2$ for quarkonium production at mid rapidity, the short distance part  can interfere with long distance part. Therefore, the $v$-expansion for the long distance matrix elements can be unclear. 

Nevertheless, in the very forward rapidity region, thanks to the Lorentz time dilation, one can find
\begin{align}
\frac{1}{m_qv}\frac{P_{\parallel}}{M}\gg \frac{1}{P_\perp}\sim\frac{1}{Q_{sA}}
\label{eq:effective-factorization}
\end{align}
where $P_{\parallel}\approx M\cosh y$. Therefore, at $y\gg \ln\frac{2m_qv}{P_\perp}\sim\ln\frac{Mv}{Q_{sA}}$, the effective factorization between the $q\bar q$ pair production and the bound state formation is justified in both CEM and NRQCD approach except for the case of $v\rightarrow0$, which corresponds to Color Singlet Model. Indeed, $v\rightarrow0$ allows Eq.~(\ref{eq:effective-factorization}) to be valid even at backward rapidity ($y\ll 0$). However, the small-$x$ formalism is not suitable to describe the $q\bar q$ pair production in p$A$ collisions at backward rapidity, as explained in the next section. We must keep in mind in this respect.

Here, a comment on the effective factorization is noted. In the above discussion, we assume that the $q\bar q$ can be transmuted into quarkonium enough outside of target nucleus without any nuclear medium effect in final state. In fact, we might need to take into consideration final state interaction such as comover interaction in the small-$x$ formalism to describe $\psi(2S)$ suppression in p$A$ collisions~\cite{Gavin:1996yd,Ferreiro:2014bia}. In general, final state interaction can violate the effective factorization. Therefore, the above discussion could not be justified for $\psi(2S)$ production. This issue is beyond the scope of this paper. Therefore, we restrict ourself to ground state such as $J/\psi$ and $\Upsilon$ by assuming that final state interaction can be negligible for their production.


\section{Heavy quark pair production in the small-$x$ formalism}\label{sec:3}

\quad
We review $q\bar q$ production in p$A$ collisions in the small-$x$ formalism. Thus far, only the $q\bar q$ production at leading order (LO) in strong coupling has been discussed in the small-$x$ formalism. In Refs.~\cite{Blaizot:2004wu,Blaizot:2004wv}, the authors derived an expression of the differential cross section for the $q\bar q$ production in p$A$ collisions in momentum space by using covariant gauge. In this expression, $k_\perp$-factorization is apparent approximately~\footnote{Exactly speaking, $k_\perp$-factorization is violated for the $q\bar q$ production in p$A$ collisions due to the multiple scattering effect. $k_\perp$-factorization is ensured when the target nucleus is dilute for which the multiple scattering of the $q\bar q$ in the target nucleus is replaced by one gluon exchange. See Refs.~\cite{Fujii:2006ab,Fujii:2005vj}.}. A similar result for single heavy quark production is derived in Ref.~\cite{Tuchin:2004rb} but the derivation has been completed in coordinate space by using light cone gauge. Furthermore, they independently improved their results in order to take into account the quantum evolution effect via the JIMWLK or the BK equation~\cite{Fujii:2006ab,Kovchegov:2006qn}. In fact, many recent numerical calculations of quarkonium production in p$A$ collisions in the small-$x$ formalism are based on Refs.~\cite{Blaizot:2004wu,Blaizot:2004wv}. Therefore, in this paper, we restrict ourself to the small-$x$ formalism derived in Ref.~\cite{Blaizot:2004wu,Blaizot:2004wv}.

\subsection{Framework}

\begin{figure}
\centering{
  \includegraphics[height=2.5cm]{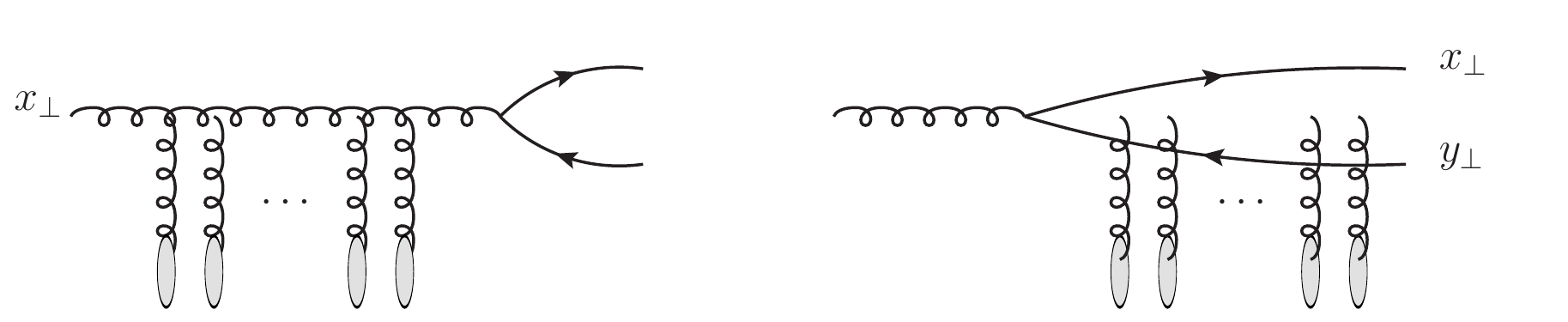}
\caption{Partonic scatterings at leading order for heavy quark pair production in p$A$ collisions.}}
\label{fig:LO}
\end{figure}

\quad
In the small-$x$ formalism, p$A$ collision can be regarded as a collision of two classical fields moving along different light cone axes. Here we choose the light cone frame in which projectile proton is going along light cone $``+"$ axis while target nucleus is moving along light cone $``-"$ axis. Valence partons inside proton and nucleus are described by solving classical Yang-Mills equation~\cite{Blaizot:2004wu}: 
\begin{align}
[D_\mu,F^{\mu\nu}]=J^\nu
\end{align}
with $J^\nu=g\delta^{\nu+}\delta(x^-)\rho_{\rm p}(x_\perp)+g\delta^{\nu-}\delta(x^+)\rho_{A}(x_\perp)$ being a color current of p$A$ collision. $\rho_{\rm p}$ ($\rho_A$) is a color charge density of valence parton in the proton (nucleus). The solution of the classical Yang-Mills equation corresponds to the background gauge field created in p$A$ collision. The order of the background gauge field in p$A$ collision is ${\cal O}(\rho^1_{\rm p}\rho_A^\infty)$. In the meantime, small-$x$ gluons are radiated from large-$x$ valence partons. This radiation process can be described by the JIMWLK equation.

 The $q\bar q$ production amplitude in p$A$ collisions in the small-$x$ formalism can be expressed as
\begin{align}
M_{s_1s_2;ij}(q,p)=&\frac{g^2}{(2\pi)^4}\int d^2k_\perp d^2k_{1\perp}\frac{\rho_{\rm p}(k_{1\perp})}{k_{1\perp}^2}\int d^2x_\perp d^2y_\perp e^{ik_\perp\cdot x_\perp}e^{i(P_\perp-k_\perp-k_{1\perp})\cdot y_\perp}\non
\times&\bar{u}_{s_1,i}\left(q\right)\left[T_g(k_{1\perp})t^bW^{ba}(x_\perp)+T_{q\bar q}(k_{1\perp},k_\perp)U(x_\perp)t^aU^\dagger(y_\perp)\right]v_{s_2,j}\left(p\right)
\label{eq:pair-amplitude}
\end{align}
where 
\begin{align}
&T_{q\bar{q}}(k_{1\perp},k_{\perp})\equiv
\frac{\gamma^+(\slq-\slk+m)\gamma^-(\slq-\slk-\slk_1+m)\gamma^+}
{2p^+[(q_\perp-k_\perp)^2+m^2]+2q^+[(q_\perp-k_\perp-k_{1\perp})^2+m^2]}, 
\notag\\
&T_{g}(k_{1\perp})\equiv
\frac{\slC_{_{L}}(p+q,k_{1\perp})}{(p+q)^2}.
\label{eq:Tqqbar-Tg}
\end{align}
$C_L^\mu(p+q,k_{1\perp})$ is the Lipatov effective vertex and its components are defined as follows:
\begin{align}
C_{_L}^+(q,k_{1\perp}) \equiv \frac{-k_{1\perp}^2}{q^-}+ q^+,~~~~
C_{_L}^-(q,k_{1\perp}) \equiv \frac{k_{2\perp}^2}{q^+}-q^-,~~~~
C_{_L}^i(q,k_{1\perp}) \equiv -2k_1^i +q^i.
\label{eq:Lipatov}
\end{align}
$u$ and $v$ are spinors of a quark and antiquark with spin $s_1$ and $s_2$, respectively. $q$ and $p$ are three momentums of a quark and antiquark, respectively. $k_{1\perp}$ is the transverse momentum of the incident gluon from the projectile proton. $k_{2\perp} \equiv p_\perp + q_\perp - k_{1\perp}$ is the transverse momentum transfer from the target nucleus to the $q\bar q$ pair carried by the gluons from the target nucleus . Eq.~(\ref{eq:pair-amplitude}) represents two physical processes as depicted in Fig.~\ref{fig:LO}: (i) an incident gluon suffers multiple scattering with many gluons in the target nucleus and subsequently splits into a $q\bar q$ pair. (ii) a $q\bar q$ pair is produced first and subsequently scatters off the target nucleus. In the high scattering energy, the transverse positions of the $q\bar q$ and gluon are almost frozen during they pass through the target nucleus. Therefore, in Eq.~(\ref{eq:pair-amplitude}), we express the multiple scattering of the quark (antiquark) and the gluon in the target nucleus by the Wilson line in the fundamental and adjoint representation with Eikonal approximation:
\begin{align}
U(x_\perp)&={\cal P}_+\exp\left[ig\int_{-\infty}^{+\infty}dz^+A^-_A(z^+,x_\perp)\cdot t\right],\notag\\
W(x_\perp)&={\cal P}_+\exp\left[ig\int_{-\infty}^{+\infty}dz^+A^-_A(z^+,x_\perp)\cdot T\right]
\end{align}
where $A^-(z^+,x_\perp)$ is the gluon field inside the target nucleus. $t^a$ and $T^a_{ij}$ are the generators of SU(3) group in the fundamental and adjoint representation, respectively.

The $q\bar q$ production cross section for minimum bias event in p$A$ collisions at classical level can be obtained by averaging the squared amplitude over the distributions of the classical color sources $\rho_p$ and $\rho_A$ as follows:	
\begin{align}
\frac{d\hat{\sigma}_{q\bar{q}}}{d^2q_{\perp} d^2p_{\perp} dy_q dy_{p}}=\frac{1}{[2(2\pi)^3]^2}\int d^2b_\perp\int{\cal D}\rho_{\rm p}{\cal D}\rho_AW_{\rm p}[\rho_{\rm p}]W_{A}[\rho_A]|M_{s_1s_2;ij}(q,p)|^2
\label{eq:qq-xsection-classical}
\end{align}
where $W_{\rm p}$ and $W_{A}$ are the weight functionals of $\rho_{\rm p}$ and $\rho_A$, respectively. $b_\perp$ is impact parameter in p$A$ collision. Rapidity dependence of Eq.~(\ref{eq:qq-xsection-classical}) is embodied in the weight functionals which obey the JIMWLK equation~\cite{Kovchegov:2012mbw}.

\subsection{Multi-point Wilson line correlators}

\quad
In general, wave function of target nucleus in high energy can be expressed as a correlation function of multi-point Wilson line. The multi-point Wilson line correlator depends on the color state of the produced $q\bar q$ pair in p$A$ collisions. Therefore, the multi-point Wilson line correlator can be much complicated in general. However, large-$N_c$ approximation allows us to simplify the multi-point Wilson line correlator. We clarify in this respect below.

\subsubsection{Summing over the color of the produced $q\bar q$}

\quad
First of all, let us consider the case that all the spin and the color of the produced $q\bar q$ pair are summed over, since it is straight forward to see how the multi-point Wilson line correlator appears in this  case. By summing over the spin and the color of the produced $q\bar q$ pair, Eq.~(\ref{eq:qq-xsection-classical}) can be cast into
\begin{align}
\frac{d \sigma_{q \bar{q}}}{d^2q_{\perp} d^2p_{\perp} dy_q dy_p}
=&
\frac{\alpha_s}{(2\pi)^6 C_F}
\int\frac{d^2k_{1\perp}}{(2\pi)^2}
\frac{\varphi_{{\rm p},Y_{\rm p}}(k_{1\perp})}{k_{1\perp}^2 k_{2\perp}^2}\non
&\times\underbrace{\left[\int\frac{d^2k_\perp d^2k_\perp^\prime}{(2\pi)^4}\Xi^{q\bar q, q\bar q} \phi^{q\bar q, q\bar q}_{A,Y_A}
+\int\frac{d^2k_\perp}{(2\pi)^2}\Xi^{q\bar q, g} \phi^{q\bar q,g}_{A,Y_A}
+\Xi^{g,g} \phi^{g,g}_{A,Y_A}
\right]}
_{\color{black}{\Longrightarrow \int\frac{d^2k_\perp}{(2\pi)^2}\Xi(k_{1\perp},k_{2\perp},k_\perp) \phi^{q\bar q,g}_{A,Y_A}}(k_{2\perp},k_\perp)}
\label{eq:no-projection}
\end{align}
where we have used the large-$N_c$ approximation and the sum rule~\footnote{Here, we refer to the identity $\int\frac{d^2k_\perp d^2k_\perp^\prime}{(2\pi)^4}\phi^{q\bar q, q\bar q}_{A,Y_A}=\int\frac{d^2k_\perp}{(2\pi)^2}\phi^{q\bar q, g}_{A,Y_A}=\phi^{g,g}_{A,Y_A}$ as the sum rule. See Ref.~\cite{Blaizot:2004wv}.}. $\Xi=\Xi^{q\bar q, q\bar q}+\Xi^{q\bar q, g}+\Xi^{g,g}$ is the partonic hard scattering part. $\varphi_{{\rm p},Y_{\rm p}}(k_\perp)$ is the unintegrated gluon distribution function (UGDF) of projectile proton~\footnote{
$\phi_{{\rm p},x}^{g,g}$ can be reduced to $\varphi_{{\rm p},x}$ at leading twist approximation. Indeed, $\phi_{{\rm p},x}^{g,g}$ is mostly used for numerical calculations instead of $\varphi_{{\rm p},x}$ because we can easily take into account the quantum evolution effect in $\phi_{{\rm p},x}^{g,g}$ via the BK equation~\cite{Fujii:2013gxa}.} and its definition is given by
\begin{align}
\varphi_{{\rm p},Y_{\rm p}}(k_\perp)=\frac{(2\pi)^2\alpha_s S_\perp}{k_\perp^2}\int d^2x_\perp e^{ik_\perp\cdot x_\perp}\langle \rho_{\rm p}(0)\rho_{\rm p}(x_\perp)\rangle_{Y_{\rm p}}.
\label{eq:phigg}
\end{align}
The multi-point Wilson line correlator of the target nucleus in the large-$N_c$ approximation can be written as
\begin{align}
\phi_{A,Y_A}^{q\bar q,g}({k}_{2\perp},k_\perp)
&\equiv S_\perp\frac{N_ck_{2\perp}^2}{2}\int d^2x_\perp d^2y_\perp e^{ik_\perp\cdot x_\perp+i(k_{2\perp}-k_\perp)\cdot y_\perp}\frac{1}{N_c^2}\langle {\rm tr}[U(x_\perp)t^aU^\dagger(y_\perp)t^bW_{ba}(0_\perp)]\rangle_{Y_A}\non
&\approx S_\perp \,\frac{N_c k_{2\perp}^2}{4}F_{Y_A}({k}_{2\perp}-k_\perp)F_{Y_A}(k_\perp)
\label{eq:phiqqg}
\end{align}
where we have used the Fierz identity as well as the identity $t^bW^{ba}(x_\perp)=U(x_\perp)t^aU^\dagger(x_\perp)$.
Fourier transform of the fundamental dipole amplitude is defined by
\begin{align}
{F}_{Y}(k_{\perp})
\equiv\int d^2x_\perp e^{-ik_{\perp}\cdot x_\perp} {S}_{Y}(x_\perp)=\int d^2x_\perp e^{-ik_{\perp}\cdot x_\perp} \frac{1}{N_c}\left<{\mathrm Tr}\left[U(x_\perp)U^\dagger(0_\perp)\right]\right>_{Y}.
\end{align}
The subscript $Y_{\rm p}=\ln1/x_{\rm p}$ of $\langle \cdots \rangle$ in Eq.~(\ref{eq:phigg}) represents the rapidity gap between the produced $q \bar q$ and the projectile proton. Similarly, $Y_{A}=\ln1/x_{A}$ in Eq.~(\ref{eq:phiqqg}) represents the rapidity gap between the produced $q \bar q$ and the target nucleus. $x_{\rm p}=k^+/P_{\rm p}^+$ and $x_A=k^-/P^{-}_A$ are longitudinal momentum fraction of the gluon inside the projectile proton and the target nucleus, respectively. By assuming that the impact parameter dependence of the $q\bar q$ production in p$A$ collisions is weak, the integral over the impact parameter converts into the transverse area of the nucleus $S_\perp$ explicitly. As we will see later, Eq.~(\ref{eq:no-projection}) is used as the $q\bar q$ production part in the CEM to compute quarkonium production cross section in p$A$ collisions.

\subsubsection{Projection into the color singlet and octet state for the produced $q\bar q$}

\quad
More complicated calculations are required to describe the $q\bar q$ production in color singlet state and color octet state in p$A$ collisions. In this respect, the use of NRQCD projection operators in association with the spin, the angular momentum, and the color of the produced $q\bar q$ pair is essential. As derived in Ref.~\cite{Kang:2013hta}, by using the NRQCD projection operators, Eq.~(\ref{eq:pair-amplitude}) can be cast into 
\begin{align}
M^{J_z,(1,8c)}(P)=&g^2\int \frac{d^2k_\perp d^2k_{1\perp}}{(2\pi)^4}\frac{\rho_{\rm p}(k_{1\perp})}{k_{1\perp}^2}\int d^2x_\perp d^2y_\perp e^{ik_\perp\cdot x_\perp}e^{i(P_\perp-k_\perp-k_{1\perp})\cdot y_\perp}\non
\times&\left\{{\rm tr}[C^{(1,8c)}t^bW^{ba}(x_\perp)]{\cal F}_g^{J_z}(P,k_{1\perp})+{\rm tr}[C^{(1,8c)}U(x_\perp)t^aU^\dagger(y_\perp)]{\cal F}^{J_z}_{q\bar q}(P,k_{1\perp},k_\perp)\right\}
\label{eq:projected-amplitude}
\end{align}
according to the spin and the color of the produced $q\bar q$ pair. Here,  
\begin{align}
{\cal F}^{J_z}_{q\bar q(g)}=&\sum_{L_z,S_z}\langle LL_z;SS_z|JJ_z\rangle 
\times\left\{
\begin{array}{c}
\left.{\rm tr}\left[\Pi^{S,S_z}T_{q\bar q(g)}\right]\right|_{l=0}~~~(\text{S-wave}) \\
\varepsilon^\ast_{\mu}(L_z) \left.\frac{\partial}{\partial l^\mu}{\rm tr}\left[\Pi^{S,S_z}T_{q\bar q(g)}\right]\right|_{l=0}~~~(\text{P-wave}).
\end{array}
\right.
\end{align}
$P$ is the total momentum of the $q\bar q$ pair and $l$ is the relative momentum between the quark and antiquark. $\varepsilon^\ast_{\mu}(L_z)$ is the polarization vector of the produced $q\bar q$ with the angular momentum $L_z$. In Ref.~\cite{Kang:2013hta}, the color projection operators are defined as $C^1=1/\sqrt{N_c}$ and $C^{8c}=\sqrt{2}t^c$ and the covariant spin projection operators are defined as
\begin{align}
\Pi^{S,S_z}=\sqrt{\frac{1}{m}}\sum_{s_1,s_2}\Big< \frac{1}{2},s_1;\frac{1}{2},s_2\Big|S,S_z\Big> v(p)\bar{u}(q)
\end{align}
where $\sqrt{\frac{1}{m}}$ is the normalization factor~\footnote{One can find in Ref.~\cite{Kang:2013hta} that $\bar u u=-\bar v v=2m$ for free Dirac spinors and $\langle q\bar q|q\bar q\rangle=4m$ for composite field yield the normalization factor $\frac{\sqrt{4m}}{\sqrt{2m}\sqrt{2m}}=\sqrt{\frac{1}{m}}$.}.

By using Eq.~(\ref{eq:projected-amplitude}), the differential cross section for the $q\bar q$ production in color singlet intermediate state in p$A$ collisions can be written as
\begin{align}
&\frac{d \sigma_{q \bar{q}}^{\rm CS}}{d^2P_{\perp}dy}
=
\frac{2\alpha_s}{(2\pi)^3(N_c^2-1)}
\int \frac{d^2k_{1\perp}d^2k_{\perp}d^2k^\prime_{\perp}}{(2\pi)^6}\frac{\varphi_{{\rm p},Y_{\rm p}}(k_{1\perp})}{k_{1\perp}^2}
\frac{1}{2J+1}\sum_{J_z}{\cal F}^{J_z}_{q\bar q}(P,k_{1\perp},k_\perp){\cal F}^{J_z\dagger}_{q\bar q}(P,k_{1\perp},k^\prime_\perp)
\non
&\times
\int d^2x_\perp d^2x^\prime_\perp d^2y_\perp d^2y^\prime_\perp
e^{i(k_\perp\cdot x_\perp- k^\prime_\perp\cdot x^\prime_\perp)}
e^{i(k_{2\perp}-k_\perp)\cdot y_\perp}
e^{-i(k_{2\perp}-k^\prime_\perp)\cdot y^\prime_\perp}
\frac{1}{N_c}\langle{\rm tr}[U(x_\perp)t^aU^\dagger(y_\perp)]
{\rm tr}[U(y^\prime_\perp)t^{a}U^\dagger(x^\prime_\perp)]\rangle_{Y_A}.
\label{eq:cs-xsection}
\end{align}
One can understand immediately that the diagram depicted in Fig.~\ref{fig:LO} (left) never contribute to the color singlet $q\bar q$ production because the gluon, which is the colored object, cannot be transmuted into the color singlet object without any gluon radiation in final state. Here, by using the Fierz identity and the large-$N_c$ approximation at the last line in Eq.~(\ref{eq:cs-xsection}), the multi-point Wilson line correlator can be written as
\begin{align}
\frac{1}{N_c}\langle{\rm tr}[U(x_\perp)t^a U^\dagger(y_\perp)]
{\rm tr}[U(y^\prime_\perp)t^{a} U^\dagger(x^\prime_\perp)]\rangle_{Y_A}
\approx \frac{1}{2}\left[
Q_{Y_A}(x_\perp,y_\perp;y^\prime_\perp,x^\prime_\perp)
-S_{Y_A}(x_\perp,y_\perp)S_{Y_A}(y^\prime_\perp,x^\prime_\perp)\right]
\label{eq:four-point-correlator}
\end{align}
where the first term in the brackets is referred to as the quadrupole amplitude which is given by
\begin{align}
Q_{Y_A}(x_\perp,y_\perp;y^\prime_\perp,x^\prime_\perp)
\equiv
{\frac{1}{N_c}}{\rm tr}\langle U(x_\perp) 
U^\dagger(x^\prime_\perp)U(y^\prime_\perp) 
U^\dagger(y_\perp)\rangle_{Y_A}.
\label{eq:quadrupole-amplitude}
\end{align}
Interestingly, the quadrupole amplitude can survive even if we take the large-$N_c$ approximation. This implies that the produced $q\bar q$ pair in the color singlet state can carry the information on the quadrupole amplitude for the target nucleus. The quadrupole amplitude never appears in Eq.~(\ref{eq:no-projection}).

Similarly, the differential cross section for the $q\bar q$ pair production in the color octet state in p$A$ collisions is given by
\begin{align}
&\frac{d \sigma_{q \bar{q}}^{\rm CO}}{d^2P_{\perp}dy}
=
\frac{2\alpha_s}{(2\pi)^3(N_c^2-1)}
\int \frac{d^2k_{1\perp}d^2k_{\perp}d^2k^\prime_{\perp}}{(2\pi)^6}\frac{\varphi_{{\rm p},Y_{\rm p}}(k_{1\perp})}{k_{1\perp}^2}
\non
&\times
\int d^2x_\perp d^2y_\perp d^2x^\prime_\perp d^2y^\prime_\perp
e^{ik_\perp\cdot x_\perp-i k^\prime_\perp\cdot x^\prime_\perp}
e^{i(P_{\perp\!}- k_\perp- k_{1\perp})\cdot y_\perp}
e^{-i(P_{\perp\!}- k^\prime_\perp- k_{1\perp})\cdot y^\prime_\perp}\non
&\times
\left[
\Xi^{\rm CO}_1{\cal W}_{Y_A}({x,y ; y^\prime,x^\prime})
+\Xi^{\rm CO}_2{\cal W}_{Y_A}({x,y ; x^\prime,x^\prime})
+\Xi^{\rm CO}_3{\cal W}_{Y_A}({x,x ; y^\prime,x^\prime})
+\Xi^{\rm CO}_4{\cal W}_{Y_A}({x,x ; x^\prime,x^\prime})
\right]
\label{eq:co-xsection}
\end{align}
where $\Xi^{\rm CO}_{1\sim4}$ are the hard scattering matrix elements and 
\begin{align}
{\cal W}_{Y_A}(x,y ; y^\prime,x^\prime)
\equiv&\frac{2}{N_c^2-1}
\langle{\rm tr}[t^cU(x_\perp)t^a U^\dagger(y_\perp)]
{\rm tr}[U(y^\prime_\perp)t^{a} U^\dagger(x^\prime_\perp)t^c]\rangle_{Y_A}\non
=&\frac{1}{2(N_c^2-1)}\Big[
\langle{\rm tr}[U(y^\prime_\perp)U^\dagger(y_\perp)]
{\rm tr}[U(x_\perp)U^\dagger(x^\prime_\perp)]\rangle_{Y_A}
-\frac{1}{N_c}\langle{\rm tr}[U(x_\perp)U^\dagger(y_\perp)
U(y^\prime_\perp)U^\dagger(x^\prime_\perp)]\rangle_{Y_A}
\non
&-\frac{1}{N_c}\langle{\rm tr}[U(x_\perp)U^\dagger(x^\prime_\perp)
U(y^\prime_\perp)U^\dagger(y_\perp)]\rangle_{Y_A}
+\frac{1}{N_c^2}\langle{\rm tr}[U(x_\perp)U^\dagger(y_\perp)]
{\rm tr}[U(y^\prime_\perp)U^\dagger(x^\prime_\perp)]\rangle_{Y_A}\Big]\non
\approx
&\frac{1}{2}
S_{Y_A}(y^\prime_\perp,y_\perp)
S_{Y_A}(x_\perp,x^\prime_\perp).
\label{eq:octet-expectation}
\end{align}
In order to derive the last line in Eq.~(\ref{eq:octet-expectation}), we have used the Fierz identity and the large-$N_c$ approximation. As a result, Eq.~(\ref{eq:co-xsection}) can be reduced to
\begin{align}
&\frac{d \sigma_{q \bar{q}}^{\rm CO}}{d^2P_{\perp}dy}
=
\frac{\alpha_sS_\perp}{(2\pi)^3(N_c^2-1)}
\int \frac{d^2k_{1\perp}d^2k_{\perp}}{(2\pi)^4}\frac{\varphi_{{\rm p},Y_{\rm p}}(k_{1\perp})}{k_{1\perp}^2}
F_{Y_A}(k_{2\perp}-k_{\perp})F_{Y_A}(k_{\perp})\Xi^{\rm CO}
\label{eq:co-xsection-simplified}
\end{align}
where $\Xi^{\rm CO}=\sum_i \Xi^{\rm CO}_i=\frac{1}{2J+1}\sum_{J_z}|{\cal F}^{J_z}_{q\bar q}+{\cal F}^{J_z}_{g}|^2$. Eq.~(\ref{eq:co-xsection-simplified}) is the same as Eq.~(\ref{eq:no-projection}) to the extent that the multi-point Wilson line correlators can be expressed in terms of the dipole amplitude only, although the hard scattering matrix elements are different. 

Eq.~(\ref{eq:cs-xsection}) and Eq.~(\ref{eq:co-xsection-simplified}) are used to calculate quarkonium production in p$A$ collisions incorporated with the NRQCD long distance matrix elements. In fact, as we will see later, a contribution of the color singlet $q\bar q$ production to $J/\psi$ total cross section in pp and p$A$ collisions at RHIC and LHC can be negligible compared to a contribution of the color octet $q\bar q$ pair production. Therefore, of particular importance is that the nuclear dependence of the $q\bar q$ production cross section can be the same both in the CEM and the NRQCD approach since the multi-point Wilson line correlator in Eq.~(\ref{eq:no-projection}) is the same as that in Eq.~(\ref{eq:co-xsection-simplified})~\cite{Kang:2013hta,Qiu:2013qka}.

\subsection{Hybrid approach for forward $q\bar q$ production}

\quad
At forward rapidity\footnote{In this paper, proton moving direction (nucleus moving direction) is defined as forward (backward) rapidity.} where $x_p\sim1$, the phase space of the gluon distribution in the projectile proton shrinks. As discussed in Refs.~\cite{Fujii:2006ab,Fujii:2013gxa,Kang:2013hta}, it is not hard to understand that all of the hard matrix elements in Eqs.~(\ref{eq:no-projection})(\ref{eq:cs-xsection})(\ref{eq:co-xsection-simplified}) are quadratic in $k_{1\perp}$ when $k_{1\perp}\rightarrow 0$. Therefore, the quadratic part $k_{1\perp}^2$ in the hard matrix elements cancels out $k_{1\perp}^2$ in the denominator when $k_{1\perp}\rightarrow 0$. Thus, we can safely replace the UGDF of the projectile proton with the collinear gluon PDF~\cite{Pumplin:2002vw} when $k_{1\perp}\rightarrow 0$, although the multi-point Wilson line correlator for the target nucleus remains without any change. A combination of the collinear gluon PDF and the multi-point Wilson line correlator is referred to as Hybrid approach in contrast to the $k_t$-factorization approach. The UGDF for the projectile proton is related to the usual collinear gluon PDF through the following definition~\cite{Fujii:2006ab,Fujii:2013gxa,Kang:2013hta}:
\begin{align}
x_pG(x_p,\mu)\equiv \frac{1}{4\pi^3}\int^{\mu^2}dk_{\perp}^2\varphi_{{\rm p},Y_p}(k_\perp)
\label{eq:ugdf-cpdf}
\end{align}
where $\mu$ is factorization scale.


\section{Quantum evolution at small-$x$}\label{sec:4}

\quad
In this section, we briefly comment on the rapidity evolution of the dipole amplitude. We also comment on the quadrupole amplitude.

\subsection{The dipole amplitude}

\quad
For phenomenological study, the rapidity evolution effect at small-$x$ is important to compare theoretical results obtained in the small-$x$ formalism with data quantitatively. Essentially, the JIMWLK equation controls the rapidity dependence of the multi-point Wilson line correlator. However, as explained in the previous section, in the large-$N_c$ approximation, the multi-point Wilson line correlator can be expressed in terms of the dipole amplitude like Eq.~(\ref{eq:phiqqg}) and Eq.~(\ref{eq:octet-expectation}). Therefore, the rapidity evolution of the dipole amplitude is particularly important at small-$x$. In practice, the rapidity evolution of the dipole amplitude is usually described by the BK equation for the reason that we can save numerical costs. The BK equation is given by
\begin{align}
-\frac{dS_{Y}({r_\perp})}{dY}
 = \int d^2 r_{1\perp} \mathcal{K}(r_\perp, r_{1\perp}) 
\left[  S_{Y}({r_\perp}) - S_{Y}({r_{1\perp}})S_{Y}({r_{2\perp}})\right]
\end{align}
with $\vec{r}_\perp = \vec{r}_{1\perp}+ \vec{r}_{2\perp}$.  $\mathcal{K}$ is the the evolution kernel. At present, the BK equation with running coupling kernel (rcBK equation)~\cite{Balitsky:2006wa} is the state of the art technology for phenomenology~\cite{Fujii:2013gxa,Ma:2014mri,Albacete:2012xq,Albacete:2014fwa} and provides the slow evolution speed.

The initial condition for the rcBK equation can be constrained by global data analysis at HERA DIS~\footnote{In fact, it is also required to constrain not only the dipole amplitude but also the one-loop coupling constant in the coordinate space 
simultaneously. See Refs.~\cite{Albacete:2009fh,Albacete:2010sy,Lappi:2013zma}.} by using the modified McLerran-Venugopalan (MV) model~\cite{McLerran:1993ni} as the initial condition at $x_0=0.01$ with the following functional form:
\begin{align}
S_{Y=Y_0}(r_\perp)=\exp\left[-\frac{(r_\perp^2Q_{s0}^2)^\gamma}{4}\ln\left(\frac{1}{|r_\perp|\Lambda}+e_c\cdot e\right)\right].
\label{eq:rcBK-initial-condition}
\end{align}
In Refs.~\cite{Albacete:2009fh,Albacete:2010sy}, they obtained the parameters set referred to as MV$^\gamma$ with $\gamma\neq1$ and $e_c=1$. Another parameters set referred to as MV$^e$ with $\gamma=1$ and $e_c\neq 1$ are obtained in Ref.~\cite{Lappi:2013zma}. Both MV$^\gamma$ and MV$^e$ parameters set can be used for the rcBK equation for the proton. When $\gamma=1$ and $e_c=1$, Eq.~(\ref{eq:rcBK-initial-condition}) recovers the quasi-classical MV model which includes multiple scattering effect.

In contrast to the initial condition for the proton, we need to make a model for the initial condition for the rcBK equation for the nucleus due to lack of precise data of $e+A$ collision. Examples of the initial condition for the nucleus are listed below.
\begin{itemize}
\item[$\bullet$] \textit{Homogeneous approximation}\\

\quad
The saturation scale of the nucleus should be proportional to the atomic mass weight $A^{1/3}$. Therefore, we should replace the initial saturation scale for the proton with the one for the nucleus in Eq.~(\ref{eq:rcBK-initial-condition}) as follows:
\begin{align}
Q_{s0,A}^2=N_{\rm coll}\;Q_{s0}^2\approx cA^{1/3}Q_{s0}^2,
\label{eq:rcBK-IC-nucleus}
\end{align}
where $N_{\rm coll}$ or $c$ is an input parameter which can be tuned by data fitting. Eq.~(\ref{eq:rcBK-IC-nucleus}) is referred to as homogeneous approximation in this paper because the gluon density in the transverse plane is assumed to be homogeneous due to the assumption that the impact parameter dependence is weak. This assumption could be a reasonable approximation as far as we restrict ourself to minimum bias events in p$A$ collisions~\cite{Fujii:2013gxa,Ma:2015sia,Fujii:2015lld}. However, if we set $c=1$ as a naive expectation, the small-$x$ formalism provides a strong nuclear suppression of $J/\psi$ production in p$A$ collisions at LHC in the forward rapidity region as we will see in Section~\ref{sec:6}~\cite{Fujii:2013gxa}. 

\quad
Although not much data are available,  in Ref.~\cite{Dusling:2009ni}, they fitted the value of $c$ in the initial condition for the dipole amplitude for the nucleus from nuclear DIS data at CERN SPS. The fitted value is roughly $c \sim 0.5$ and it can provide a reasonable result of $J/\psi$ suppression in p$A$ collisions compared to the result with $c=1$. Of course, the small value of $c$ corresponds to the smaller value of $N_{\rm coll}$ than what we have expected. However, the physics behind it is poorly understood now.
\end{itemize}
\begin{itemize}
\item[$\bullet$] \textit{Glauber model approach}\\

\quad
In Refs.~\cite{Ducloue:2015gfa,Lappi:2013zma,Ducloue:2016pqr}, they addressed the initial condition for the nucleus by using optical Glauber model with MV$^e$ parameters set. This is because we do not need to tune $N_{\rm coll}$ when we use the Glauber approach in contrast to Eq.~(\ref{eq:rcBK-IC-nucleus}). In Refs.~\cite{Ducloue:2015gfa,Lappi:2013zma,Ducloue:2016pqr}, they set the rcBK initial condition for the nucleus to be
\begin{align}
S_{Y=Y_0}(r_\perp;b_\perp)=\exp\left[-AT_A(b_\perp)\frac{\sigma_0}{2}\frac{r_\perp^2Q_{s0}^2}{4}\ln\left(\frac{1}{|r_\perp|\Lambda}+e_c\cdot e\right)\right]
\label{eq:rcBK-IC-Glauber}
\end{align}
where $T_A(b_\perp)$ is Woods-Saxon nucleon distribution. $\frac{\sigma_0}{2}=\int d^2b_\perp$ is the effective transverse area of the proton and determined by DIS data fitting~\cite{Lappi:2013zma}. In fact, the fitted value of $\frac{\sigma_0}{2}$ is smaller than $\sigma_{\rm pp}^{inel}$ which has been used in previous phenomenological study using Monte Carlo Glauber~\cite{Albacete:2012xq}. Therefore, we can estimate that the $N_{\rm coll}$ used in Refs.~\cite{Ducloue:2015gfa,Lappi:2013zma,Ducloue:2016pqr} is smaller than the one obtained in Ref.~\cite{Albacete:2012xq} due to $\frac{\sigma_0}{2}<\sigma_{\rm pp}^{inel}$. As we will see in Section~\ref{sec:6}, the Galuber model approach also provides a reasonable nuclear suppression for forward $J/\psi$ production in p$A$ collisions. The small value of $N_{\rm coll}$ in the Glauber approach corresponds to the case of the small value of $c$ in the homogeneous approximation.
\end{itemize}

Finally, we just comment on the GBW model. The use of the GBW model also allows us to take into account the rapidity evolution of the dipole amplitude easily~\cite{Kopeliovich:2001ee,Kopeliovich:2010nw,Kopeliovich:2011zz}. The dipole amplitude in the GBW model~\cite{GolecBiernat:1998js} is given by
\begin{align}
S_Y(r_\perp)=\exp\left[-\frac{Q_s^2r_\perp^2}{4}\right]
\label{eq:GBW}
\end{align}
where the saturation scale is $Q_s^2(x)=Q_{s0}^2\left(x_0/x\right)^\lambda$ with $x_0=0.000304$, $\lambda=0.288$, and $Q_{s0}^2=1\;{\rm GeV}^2$. These parameters are fitted by DIS data analysis. Therefore, the dipole amplitude can be used for the proton. The rapidity dependence of the dipole amplitude is only embedded in the saturation scale. When we apply the GBW model to the nucleus, we should replace $Q_{s0}^2$ with Eq.~(\ref{eq:rcBK-IC-nucleus}).

\subsection{Quadrupole amplitude}

\quad
The $q\bar q$ pair production in the color singlet state involves the quadrupole amplitude even if we take the large-$N_c$ approximation. The rapidity evolution of the quadrupole amplitude must be described by the JIMWLK equation although solving the JIMLWK equation numerically is more complicated than the BK equation. Interestingly, if the correlators of the color charges in the nucleus are Gaussian forms, the quadrupole amplitude can be expressed in terms of the dipole amplitude only. For example, the specific expressions of the dipole amplitude is given as follows~\cite{Blaizot:2004wv,Dominguez:2011wm}:
\begin{align}
Q_{Y}(x_\perp,y_\perp;y^\prime_\perp,x^\prime_\perp)
\approx
S_{Y}(x_\perp,x^\prime_\perp)S_{Y}(y^\prime_\perp,y_\perp)
-&\frac{\ln\left[S_{Y}(x_\perp,y^\prime_\perp)S_{Y}(x^\prime_\perp,y_\perp)\right]
-\ln\left[S_{Y}(x_\perp,{y}_\perp)S_{Y}(x^\prime_\perp,y^\prime_\perp)\right]}
{\ln\left[S_{Y}(x_\perp,x^\prime_\perp)S_{Y}(y^\prime_\perp,y_\perp)\right]
-\ln\left[S_{Y}(x_\perp,{y}_\perp)S_{Y}(x^\prime_\perp,y^\prime_\perp)\right]}
\non
&\times\left[S_{Y}(x_\perp,x^\prime_\perp)S_{Y}(y^\prime_\perp,y_\perp)
-S_{Y}(x_\perp,{y}_\perp)S_{Y}(x^\prime_\perp,y^\prime_\perp)\right]
\label{eq;quadrupole}
\end{align}
where the large-$N_c$ approximation has been assumed. In this case, the rapidity dependence of the quadrupole amplitude can be calculated in terms of the rapidity evolution of the dipole amplitude. As reported in Ref.~\cite{Dumitru:2011vk}, the Gaussian approximation can provide a good description of the rapidity dependence of the quadrupole amplitude compared to the direct numerical results of the JIMWLK equation for the quadrupole amplitude. In Ref.~\cite{Ma:2014mri}, a different form of the quadrupole amplitude is proposed in the Gaussian approximation.


\section{Quarkonium production models}\label{sec:5}

\quad
In this section, we review the concepts of CEM, CSM, and NRQCD approach for describing bound state formation. These specific models have been incorporated with the small-$x$ formalism to describe $P_\perp$ distribution of $J/\psi$ and $\Upsilon$ production in high energy p$A$ collisions.

\subsection{Color Evaporation Model}

\quad
Color Evaporation Model (CEM) is a simple phenomenological model based on quark-hadron duality and has succeeded in describing many data~\cite{Brambilla:2004wf,Brambilla:2010cs}. In the CEM, differential cross section for quarkonium production can be written as
\begin{align}
\frac{d\sigma_\psi}{d^2P_\perp dY}=F_{q\bar q\rightarrow \psi}\int_{2m_q}^{2M_h}dM\frac{d\sigma_{q\bar q}}{dM d^2P_\perp dY}.
\end{align}
The produced $q\bar q$ pair is going to be bound into a quarkonium $\psi$ with the probability $F_{q\bar q\rightarrow \psi}$. Indeed, $F_{q\bar q\rightarrow \psi}$ is a universal empirical factor and should be interpreted as a normalization factor for inclusive $\psi$ production. One must keep in mind that any $K$-factor in association with higher order correction is included in $F_{q\bar q\rightarrow \psi}$. 

Since the CEM allows us to perform numerical calculations easily, the previous phenomenological studies in the small-$x$ formalism have employed the CEM~\cite{Fujii:2013gxa,Ducloue:2015gfa,Fujii:2006ab,Fujii:2015lld,Ducloue:2016pqr}.
The $q\bar q$ pair production cross section $d\sigma_{q\bar q}$ is computed by using Eq.~(\ref{eq:no-projection}). In the CEM, all the $q\bar q$ pairs transmute into quarkonium with the same transition probability. Therefore, the color octet $q\bar q$ pair is a dominant channel. The dominance of the color octet channels corresponds to the concept of the large-$N_c$ approximation.

\subsection{Color Singlet Model}

\quad
In Color Singlet Model (CSM), a $q\bar q$ pair with right quantum number is only considered in contrast to the CEM. The CSM has been studied actively after the discovery of $J/\psi$. The advantage of the CSM is that we do not need any parameter in association with the bound state formation. However, it is known that the CSM in the collinear factorization framework cannot describe $J/\psi$ production at high $P_\perp$ in hadronic collisions at collider energies~\cite{Lansberg:2008gk}. 

Nevertheless, the CSM has been employed in the many previous papers~\cite{Kharzeev:2005zr,Kopeliovich:2001ee,Kopeliovich:2010nw,Kopeliovich:2011zz,Dominguez:2011cy,Kharzeev:2012py,Kharzeev:2008nw}, although the effective factorization is unclear for the CSM as explained in the previous section. One may expect that the small-$x$ formalism incorporated with the CSM at LO does not contribution to $J/\psi$ production in high energy p$A$ collisions. However, interestingly, as indicated in Refs.~\cite{Kharzeev:2005zr,Kharzeev:2008nw}, color singlet direct quarkonium production can be enhanced in p$A$ collisions compared to the production in pp collisions. 

To see the enhancement mechanism, let us first consider direct $J/\psi$ production in pp collisions. The quantum numbers of $J/\psi$ are $\rm J^{PC}=1^{--}$. If we assume for simply that both the protons are dilute objects, the partonic scattering process at LO in pp collisions is $g+g\rightarrow J/\psi+g$ where a soft gluon is radiated in final state. Therefore, the LO cross section for $J/\psi$ production in pp collisions is of order ${\cal O}(\alpha_s^4)$ except for the contribution from the projectile proton\footnote{
	In pp collision, the splitting process $g\rightarrow q\bar q$ is order ${\cal O}(\alpha_s)$, the gluon radiation in final state is order ${\cal O}(\alpha_s)$, the gluon exchange between $q$ or $\bar q$ and the target proton is order ${\cal O}(\alpha_s^2)$. Therefore, the LO contribution in pp collision is order ${\cal O}(\alpha_s^4)$ in total. In p$A$ collision, the multiple gluon exchange between $q$ or $\bar q$ and the target nucleus is order ${\cal O}(\alpha_s^2A^{1/3})$. Therefore, one finds the LO contribution in p$A$ collision is of order ${\cal O}(\alpha_s^5A^{2/3})$ in total. The power counting obtained in this paper is slightly different from the result in Refs.~\cite{Kharzeev:2005zr,Kharzeev:2008nw}.}.

On the other hand, in p$A$ collisions, three gluons fusion process as $g+g+g\rightarrow J/\psi$ can be significant due to the multiple scattering of the $q\bar q$ in the target nucleus. The LO cross section for $J/\psi$ production in p$A$ collision is order ${\cal O}(\alpha_s^5A^{2/3})$. Therefore, the three gluon fusion process seems to be suppressed compared to the two gluon fusion in terms of the power counting. However, the quasi-classical approximation $\alpha_s^2A^{1/3}\sim{\cal O}(1)$ leads to ${\cal O}(\alpha_s^5A^{2/3})\rightarrow {\cal O}(\alpha_s)$. Thus, the color singlet direct $J/\psi$ production can be enhanced in p$A$ collisions compared to the production in pp collisions. 

Following the above qualitative discussion, we should consider the color singlet direct $J/\psi$ production in p$A$ collisions carefully to compare the small-$x$ formalism with data. We will comment on numerical results in this respect in the next section.

\subsection{Non-Relativistic QCD}

\quad
Non-Relativistic QCD (NRQCD) factorization approach allows us to evaluate a relative contribution of the $q\bar q$ production in specific quantum state systematically~\cite{Bodwin:1994jh}. At present, the collinear factorization framework with the NRQCD long distance matrix elements (LDMEs) can describe $P_\perp$ spectrum of $J/\psi$ and $\Upsilon$ production at high $P_\perp$ in pp collisions at collider energies, although spin polarization puzzle is not resolved yet~\cite{Brambilla:2004wf,Brambilla:2010cs}.

Incorporating the NRQCD LDMEs into the small-$x$ formalism has been performed in Refs.~\cite{Ma:2015sia,Kang:2013hta,Ma:2014mri}. As discussed in Section~\ref{sec:2}, the $v$-expansion for the LDMEs is effectively ensured at forward rapidity. Therefore, quarkonium $\psi$ production in p$A$ collisions in the small-$x$ formalism incorporated with the NRQCD LDMEs can be written as 
\begin{align}
\frac{d\sigma_\psi}{d^2P_\perp dY}=\sum_{\kappa} \frac{d\hat{\sigma}_{q\bar q}^\kappa}{d^2P_\perp dY} \left<{\cal O}^\kappa_\psi\right>
\end{align}
where $d\hat{\sigma}^\kappa$ is the differential cross section for the $q\bar q$ production in quantum state $\kappa$. $\left<{\cal O}^\kappa_\psi\right>$ is referred to as the universal NRQCD LDMEs. The LDMEs for the color octet channels can be extracted by data fitting using the NRQCD collinear factorization framework at NLO~\cite{Ma:2010yw,Butenschoen:2010rq,Chao:2012iv}. The LDME for the color singlet channel is given by nonrelativistic wave function of the $q\bar q$ at origin ($v=0$).
In terms of the NRQCD power counting, the important intermediate states for $J/\psi$ and $\Upsilon$ production are listed below:
\begin{align}
^3S_1^{[1]},~~~^1S_0^{[8]},~~~^3S_1^{[8]},~~~^3P_0^{[8]}
\end{align}
where we use the standard spectroscopic notation ($^{2S+1}L^{[c]}_J$). Here, $S$ is spin, $L$ is angular momentum, $J$ is total angular momentum, and $c$ is color of the $q\bar q$ pair. For $^3S_1^{[1]}$ channel, Eq.~(\ref{eq:cs-xsection}) is used as the $q\bar q$ pair production cross section, while Eq.~(\ref{eq:co-xsection}) is used for the other color octet channels.

The NRQCD factorization approach coincides with the CSM by taking the limit $v\rightarrow 0$. As derived in Ref.~\cite{Kang:2013hta}, the color singlet channel in the small-$x$ formalism with the NRQCD LDMEs corresponds to the early results of the CSM obtained in Refs.~\cite{Dominguez:2011cy,Kharzeev:2012py} by assuming the Gaussian approximation for the quadrupole amplitude.


\section{Numerical results}\label{sec:6}

\quad
In this section, we show selected numerical results of $J/\psi$ production in the small-$x$ formalism.  We restrict ourself to forward quarkonium production in pp and p$A$ collisions in the RHIC and LHC energies.

\subsection{$P_\perp$-spectrums}

\begin{figure}
\centering
  \includegraphics[width=7.5cm]{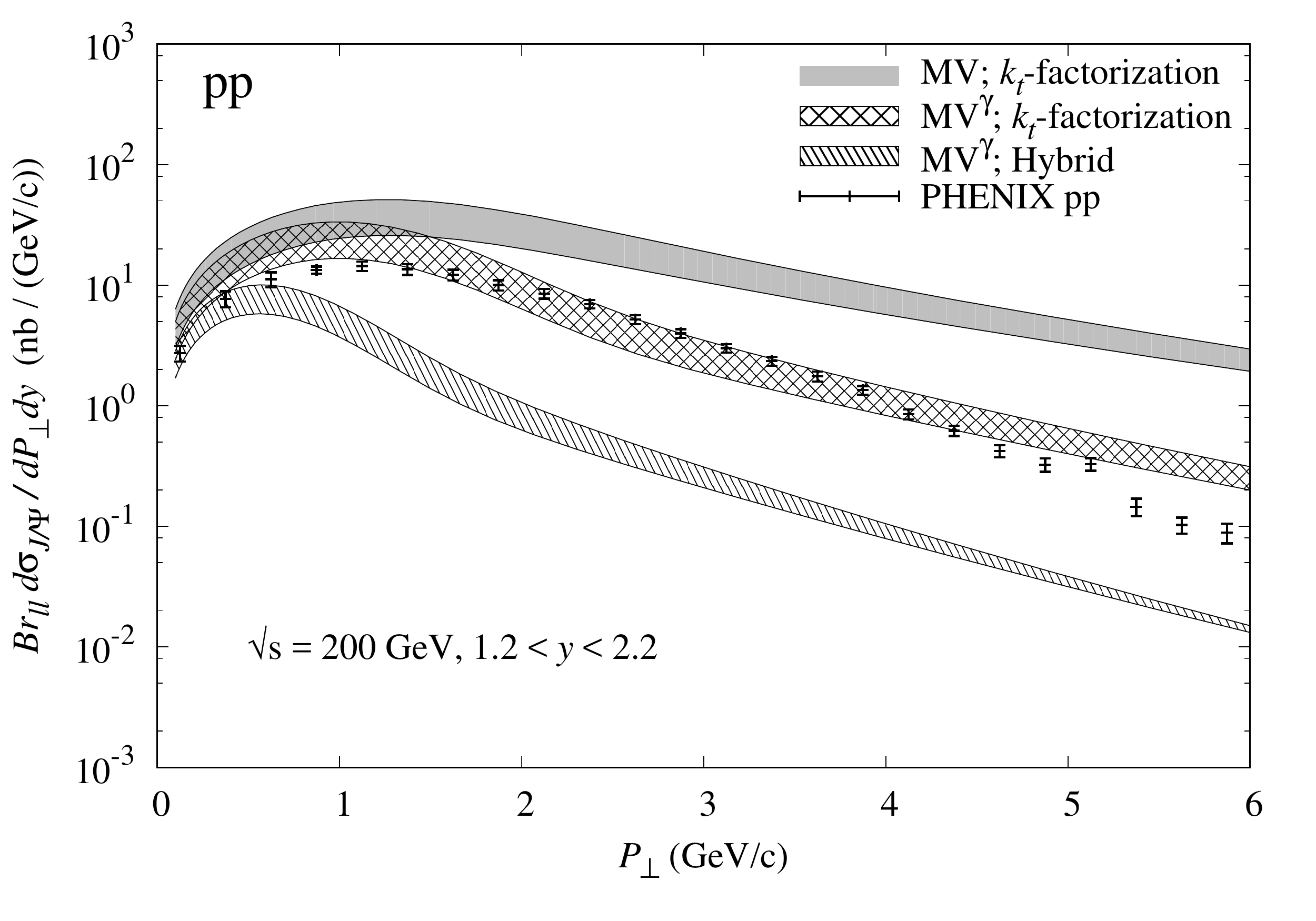}
  \includegraphics[width=7.5cm]{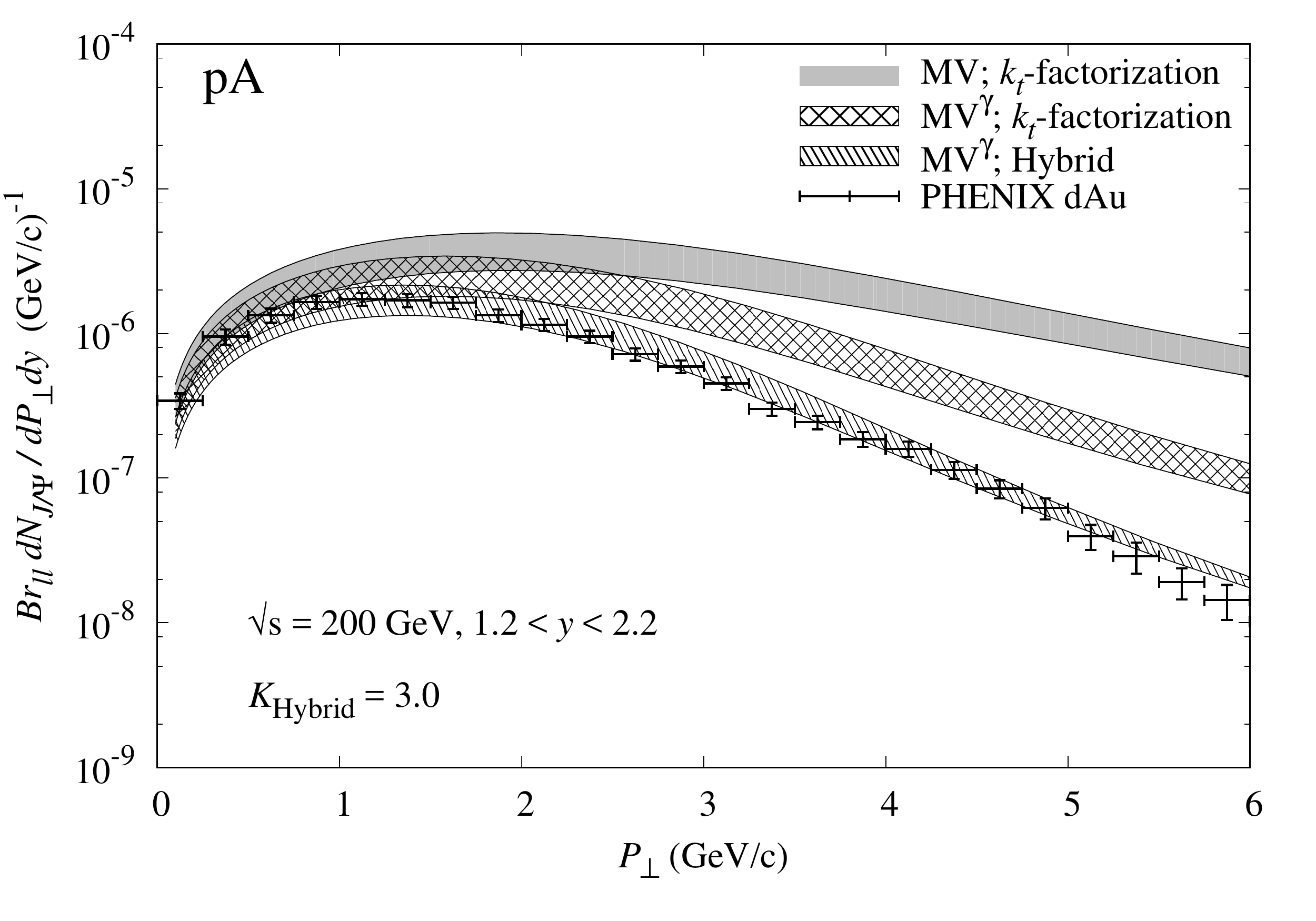}
\caption{Differential cross section as a function of $P_\perp$ for $J/\psi$ production in pp (left) and p$A$ (right) collisions at RHIC at forward rapidity. All the results are computed in the small-$x$ formalism incorporated with the CEM~\cite{Fujii:2013gxa}. 
}
\label{fig:CEM-Pt-RHIC}
\end{figure}

\begin{figure}
\centering
  \includegraphics[width=7.5cm]{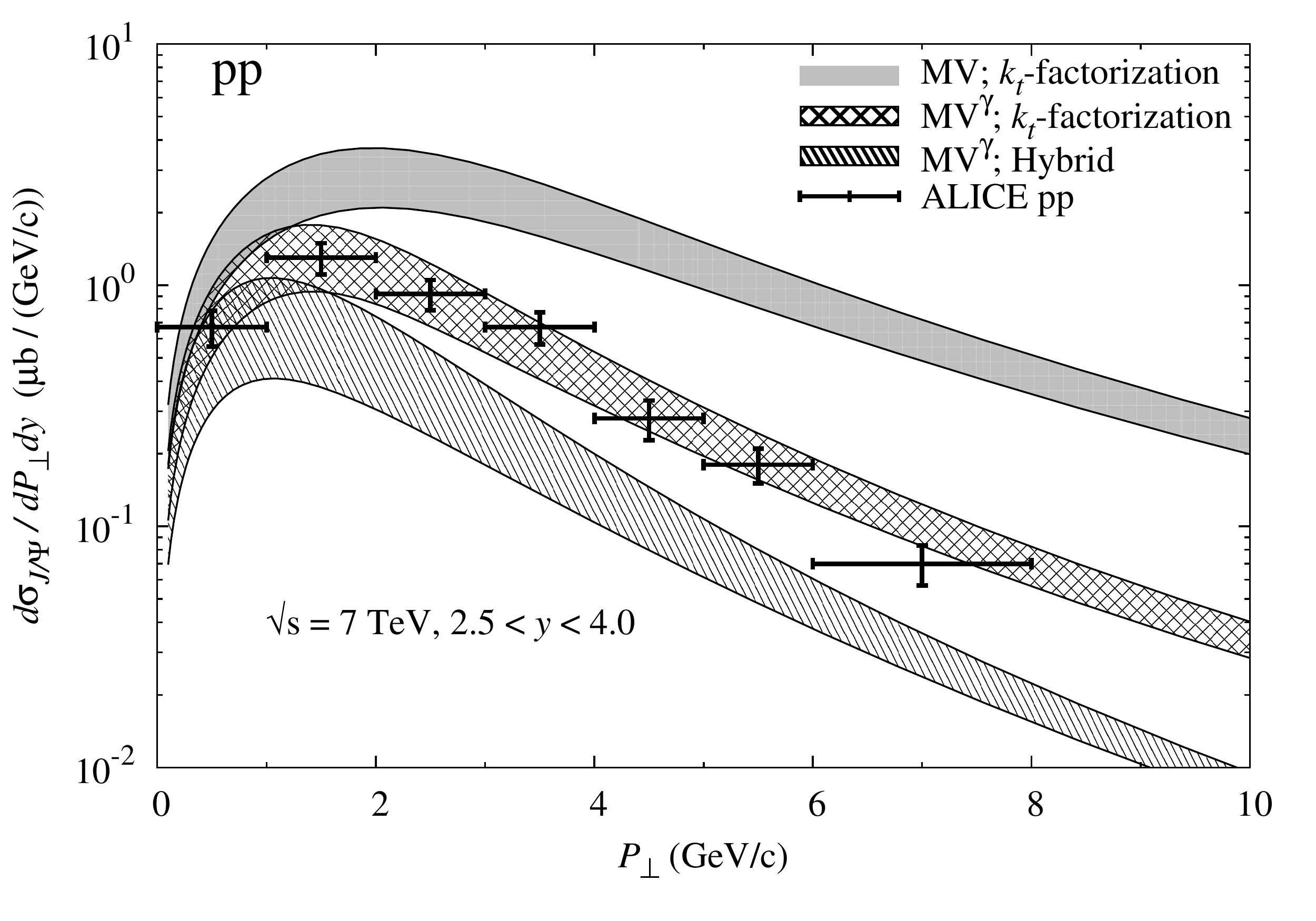}
  \includegraphics[width=7.5cm]{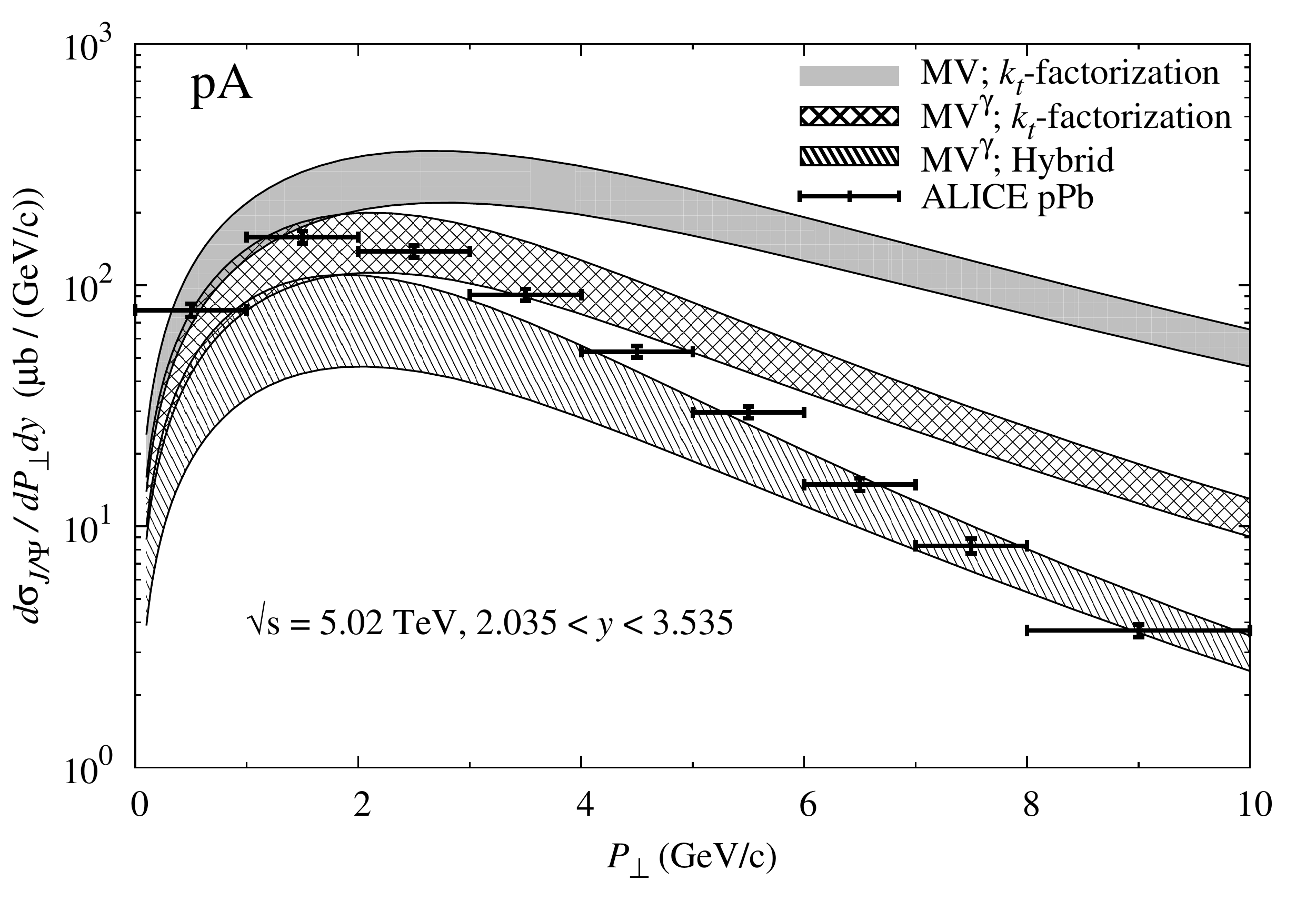}
\caption{$P_\perp$ distribution of $J/\psi$ production in pp and p$A$ collisions at LHC in the small-$x$ formalism with the CEM~\cite{Fujii:2013gxa}.}
\label{fig:CEM-Pt-LHC}
\end{figure}

\begin{figure}
\centering
  \includegraphics[width=7.5cm]{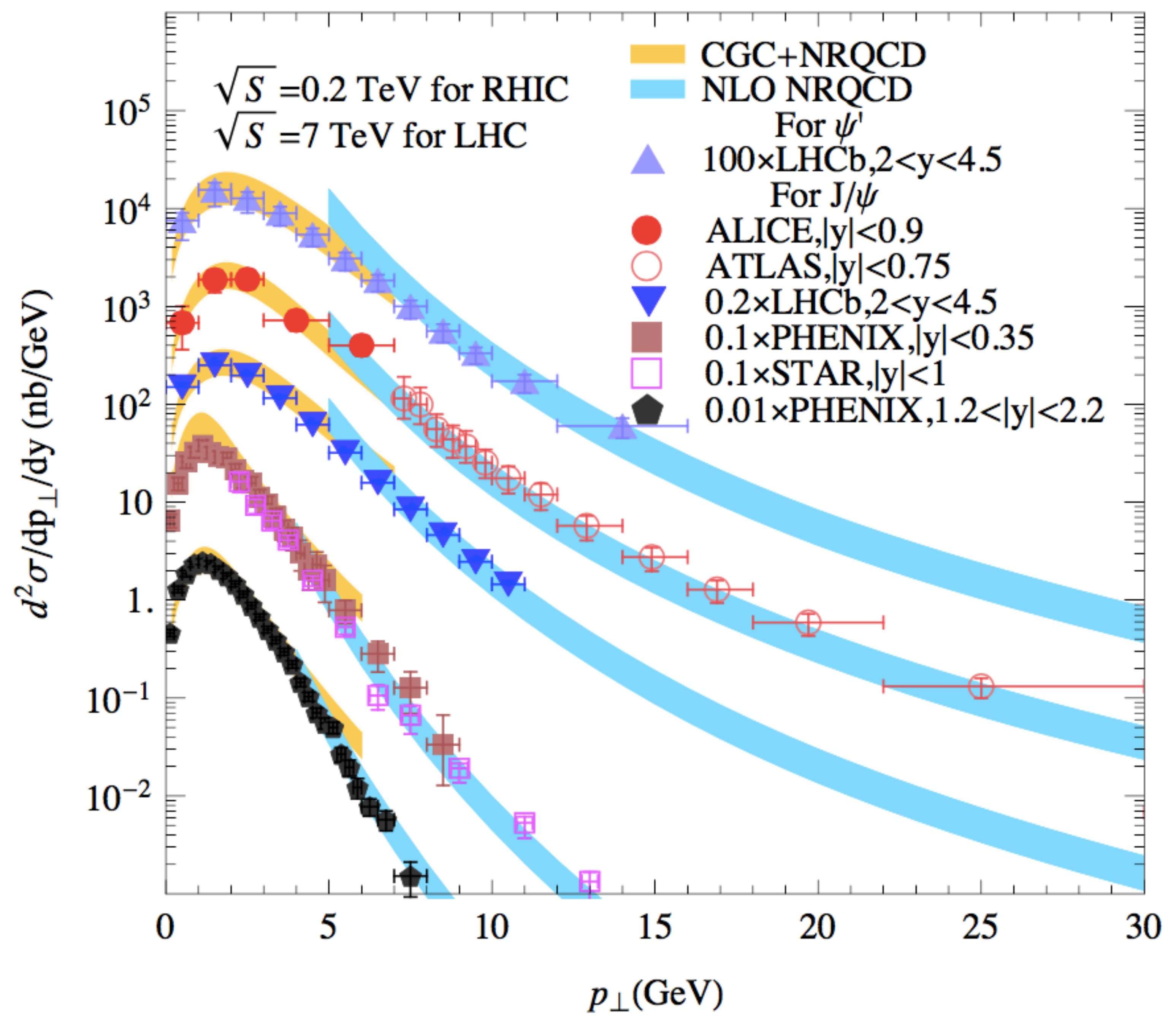}
  \includegraphics[width=7.cm]{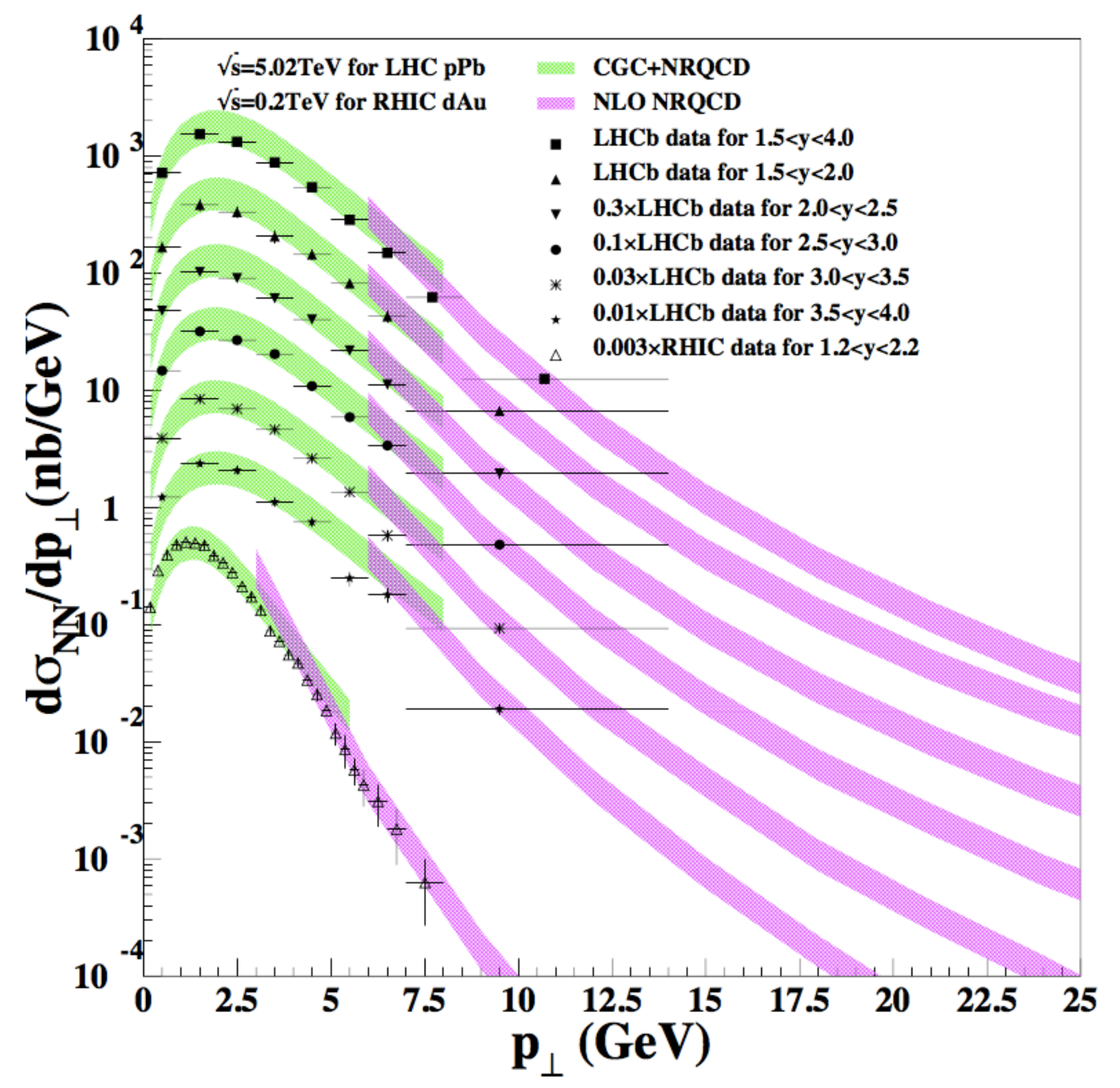}
\caption{$P_\perp$-spectrums of $J/\psi$ production in pp (left) and p$A$ (right) collisions in the small-$x$ formalism incorporated with the NRQCD LDMEs~\cite{Ma:2014mri,Ma:2015sia}. The numerical results in the NRQCD collinear factorization framework at NLO are also overlaid at higher $P_\perp$ for comparison.}
\label{fig:NRQCD-Pt}
\end{figure}

\quad
The numerical results of the differential cross section as a function of $P_\perp$ for $J/\psi$ production in pp collision at RHIC at forward rapidity are shown in Fig.~\ref{fig:CEM-Pt-RHIC} (left). The theoretical results are computed in the $k_\perp$-factorization formula and the Hybrid formula incorporated with the CEM~\cite{Fujii:2013gxa}. 
The $k_\perp$-factorized formula works moderately at lower $P_\perp\lesssim 1\;{\rm GeV}$ whereas the Hybrid formula cannot describe data even in the low $P_\perp$, since the $P_\perp$ of $J/\psi$ is only supplied by the target proton in the LO Hybrid formula although the saturation scale of the target proton is not large at RHIC. In fact, the $P_\perp$ distribution of $J/\psi$ depends on the rcBK initial condition (MV or MV$^\gamma$), since $x_A$ lies in the vicinity of $x_0=0.01$. However, we should keep in mind that $J/\psi$ production at high $P_\perp$ must be described in the collinear factorization framework rather than the small-$x$ formalism. Roughly speaking, the small-$x$ formalism can describe $J/\psi$ production in pp collisions at $P_\perp\lesssim Q_{sp}$.

We also show in Fig.~\ref{fig:CEM-Pt-RHIC} (right) the numerical results of $J/\psi$ production for minimum bias events in p$A$ collisions in the small-$x$ formalism with the CEM. The results are computed by using Eq.~(\ref{eq:rcBK-IC-nucleus}) with $c=1$ for the initial condition for the target nucleus. Due to the large saturation scale for the nucleus, the $P_\perp$ distribution of $J/\psi$ at lower $P_\perp$ can be described well by the small-$x$ formalism, although the uncertainties in association with the initial condition for the rcBK equation for the nucleus remain.

Next, we show in Fig.~\ref{fig:CEM-Pt-LHC} the results of $J/\psi$ production at LHC in the small-$x$ formalism with the CEM. Thanks to the rapidity evolution of the dipole amplitude, the uncertainty in association with the initial condition for the rcBK equation becomes slightly smaller both in pp and p$A$ collisions.

A caution concerning the UGDF of the projectile proton is noted here. In fact, an extrapolation of the UGDF at $x_p>x_0$ is required to compute the differential cross section for the forward $J/\psi$ production. For example, we can introduce a scaling function as employed in Refs.~\cite{Fujii:2013gxa,Fujii:2006ab,Fujii:2015lld}. As another choice, in Ref.~\cite{Ma:2014mri}, they consider a matching between the UGDF and the collinear gluon distribution function numerically. In this respect, one must keep in mind that the extrapolation of the UGDF at $x\geq x_0$ necessarily involves large systematic uncertainties.

The small-$x$ formalism incorporated with the CEM can moderately describe the low $P_\perp$ $J/\psi$ production in pp and p$A$ collisions at RHIC and LHC. However, in order to describe the $P_\perp$ distribution of $J/\psi$ production more correctly, we should consider the dynamics of bound state formation because we need to clarify dominant channels in $J/\psi$ production. Therefore, we must consider the small-$x$ formalism incorporated with the NRQCD LDMEs for describing $J/\psi$ production.

Fig.~\ref{fig:NRQCD-Pt} (left) shows the first numerical results for $J/\psi$ production (and also $\psi^\prime$ for comparison) in pp collisions in the small-$x$ formalism with the NRQCD LDMEs at RHIC and LHC. In Ref.~\cite{Ma:2015sia,Ma:2014mri}, they use the $k_\perp$-factorized framework with the MV initial condition for the rcBK equation. Regarding the NRQCD LDMEs, they extract those from Tevatron data of prompt $J/\psi$ production at high $P_\perp$. In Fig.~\ref{fig:NRQCD-Pt}, the results of the NRQCD approach in the usual collinear factorization framework at NLO are also shown for comparison.  One finds immediately that the small-$x$ formalism with the NRQCD LDMEs can describe all the data of $J/\psi$ production at low $P_\perp$ within the uncertainties, which are inherited from the LDMEs. On the other hand, the NRQCD collinear factorization framework can reproduce the data at larger $P_\perp$. These results suggest that we should switch from the small-$x$ formalism to the usual collinear factorization framework around $P_\perp\sim5\;{\rm GeV}$, although the quantitative value of the switching point can vary more or less. We should keep in mind that the results at mid rapidity are computed by assuming the effective factorization as explained in Section~\ref{sec:2}.

Ref.~\cite{Ma:2014mri} also reported that the contribution of the color singlet direct $J/\psi$ production in the small-$x$ formalism with the NRQCD LDMEs is about 10\% of the total cross section in pp collision at most, although the factorization for the CSM is not clear. This implies that the color singlet channel is negligible for $J/\psi$ production at low $P_\perp$ at collider energies. In other words, the color octet channels dominate in $J/\psi$ production from low $P_\perp$ to high $P_\perp$ in hadronic collisions.

The numerical results of the $P_\perp$ distribution of $J/\psi$ production in p$A$ collisions in the small-$x$ formalism with the NRQCD LDMEs are shown in Fig.~\ref{fig:NRQCD-Pt} (right). For comparison, the numerical results obtained in the collinear factorization NRQCD framework at NLO with use of nuclear PDF are also shown. One can find all the numerical results are in good agreement with data at RHIC and LHC. As mentioned in Section~\ref{sec:5}, it is expected that the color singlet channel for $J/\psi$ production can be enhanced in p$A$ collision compared to pp collision. In fact, in Ref.~\cite{Ma:2015sia}, they found that the contribution of the color singlet direct $J/\psi$ production is about 15\%-20\% at low $P_\perp$ in p$A$ collision. This value is slightly larger than that in pp collisions. However, we can conclude that the color singlet channel in the small-$x$ formalism incorporated with the NRQCD LDMEs is not important for $J/\psi$ production in both pp and p$A$ collisions after all.

\subsection{Nuclear suppression in minimum bias events}

\begin{figure}
\centering
  \includegraphics[width=7.5cm]{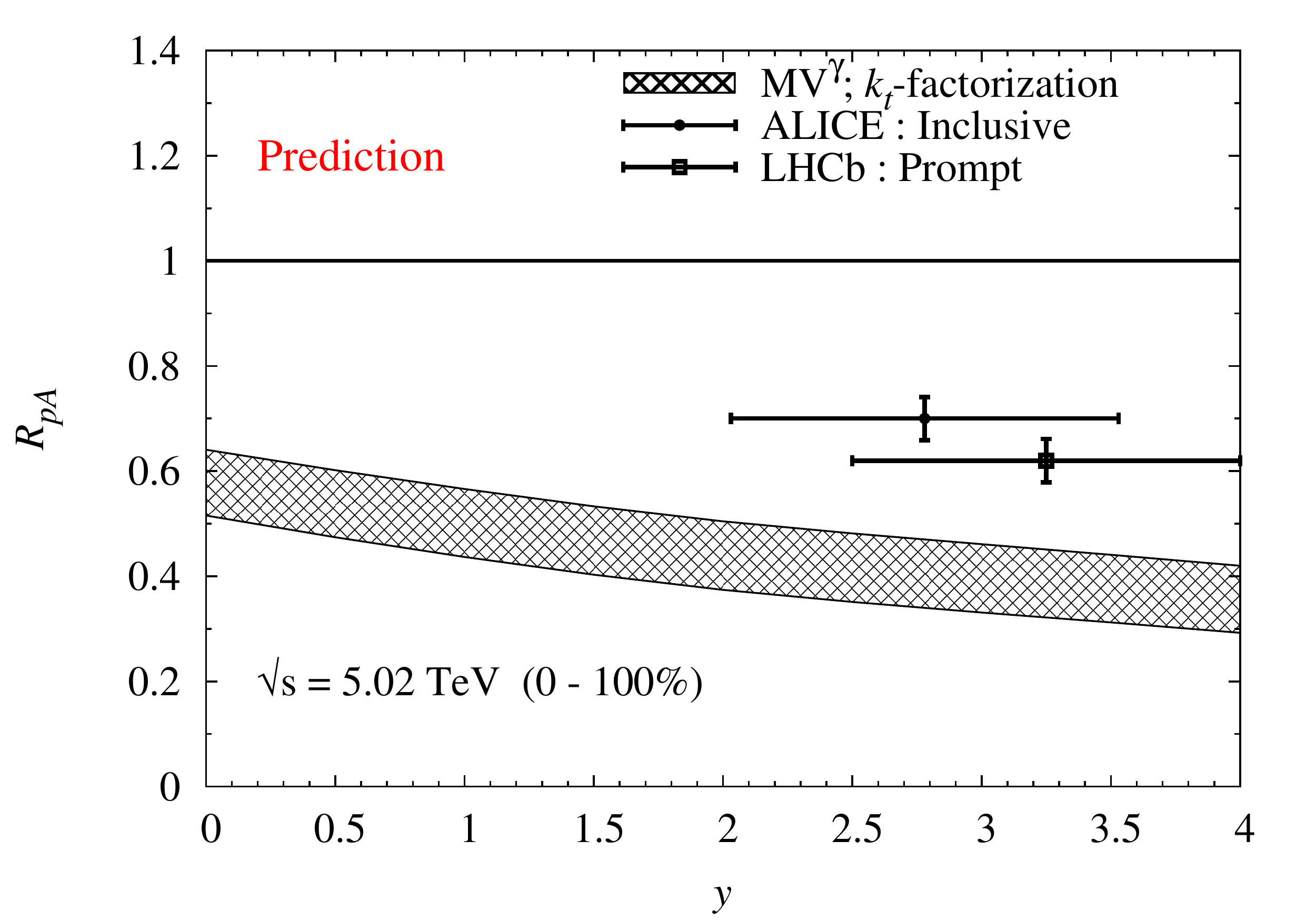}
  \includegraphics[width=7.5cm]{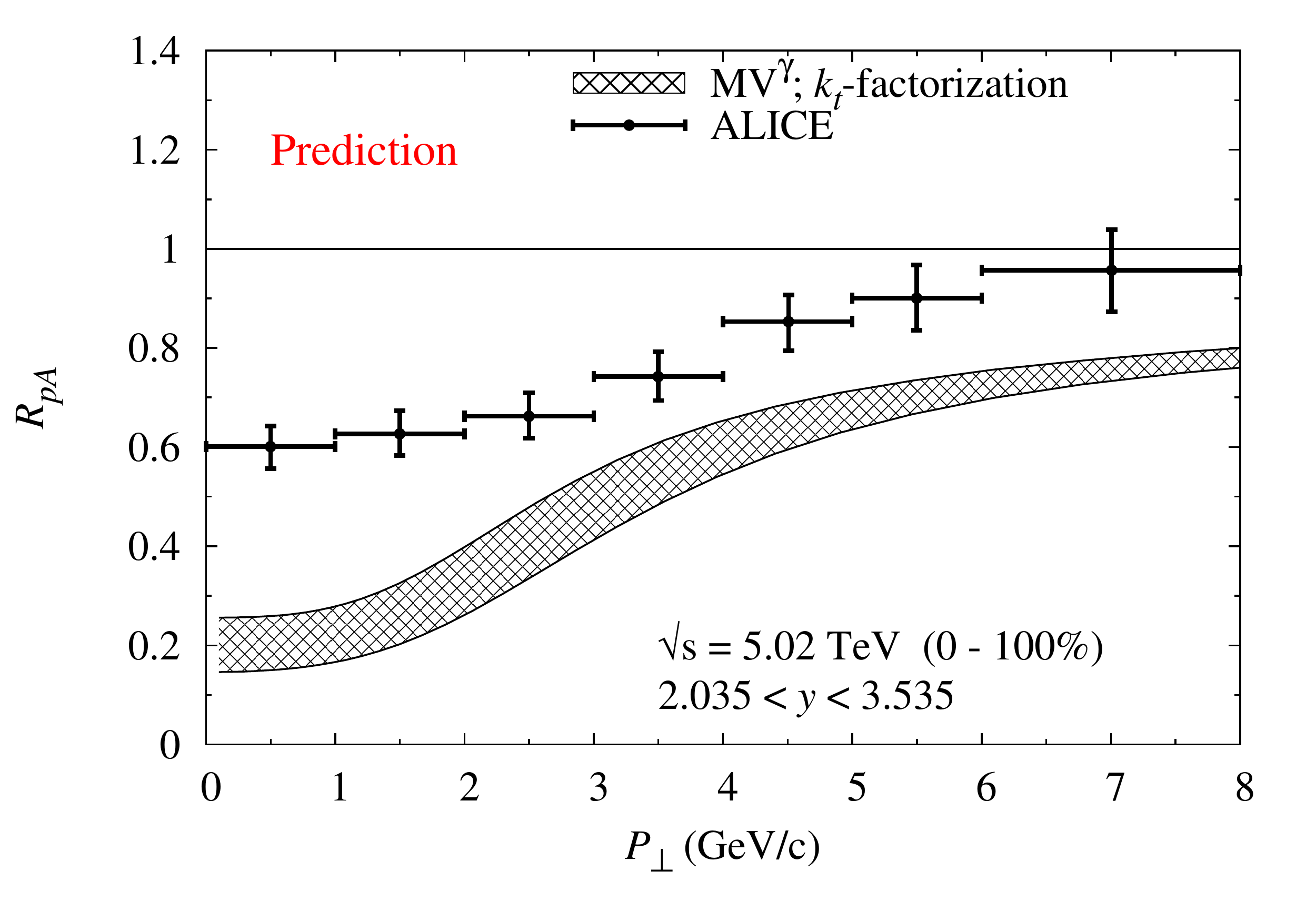}
\caption{Early predictions of $R_{{\rm p}A}$ of $J/\psi$ production at LHC in the small-$x$ formalism incorporated with the CEM by setting $c=1$ in Eq.~(\ref{eq:rcBK-IC-nucleus})~\cite{Fujii:2013gxa}. }
\label{fig:RpA-LHC-prediction}
\end{figure}

\begin{figure}
\centering
  \hspace{0.25cm}
  \includegraphics[width=7.cm]{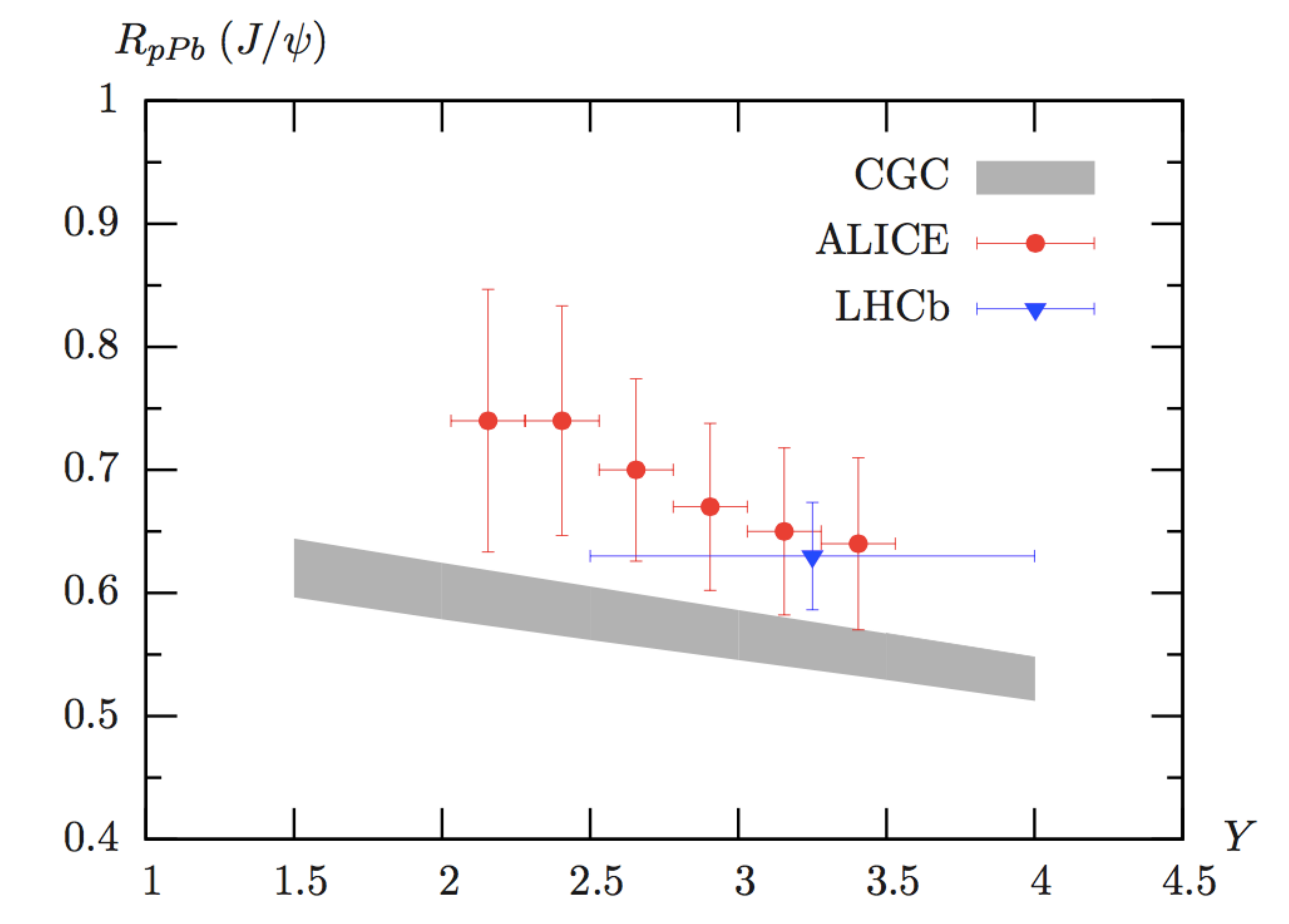}
  \includegraphics[width=7.5cm]{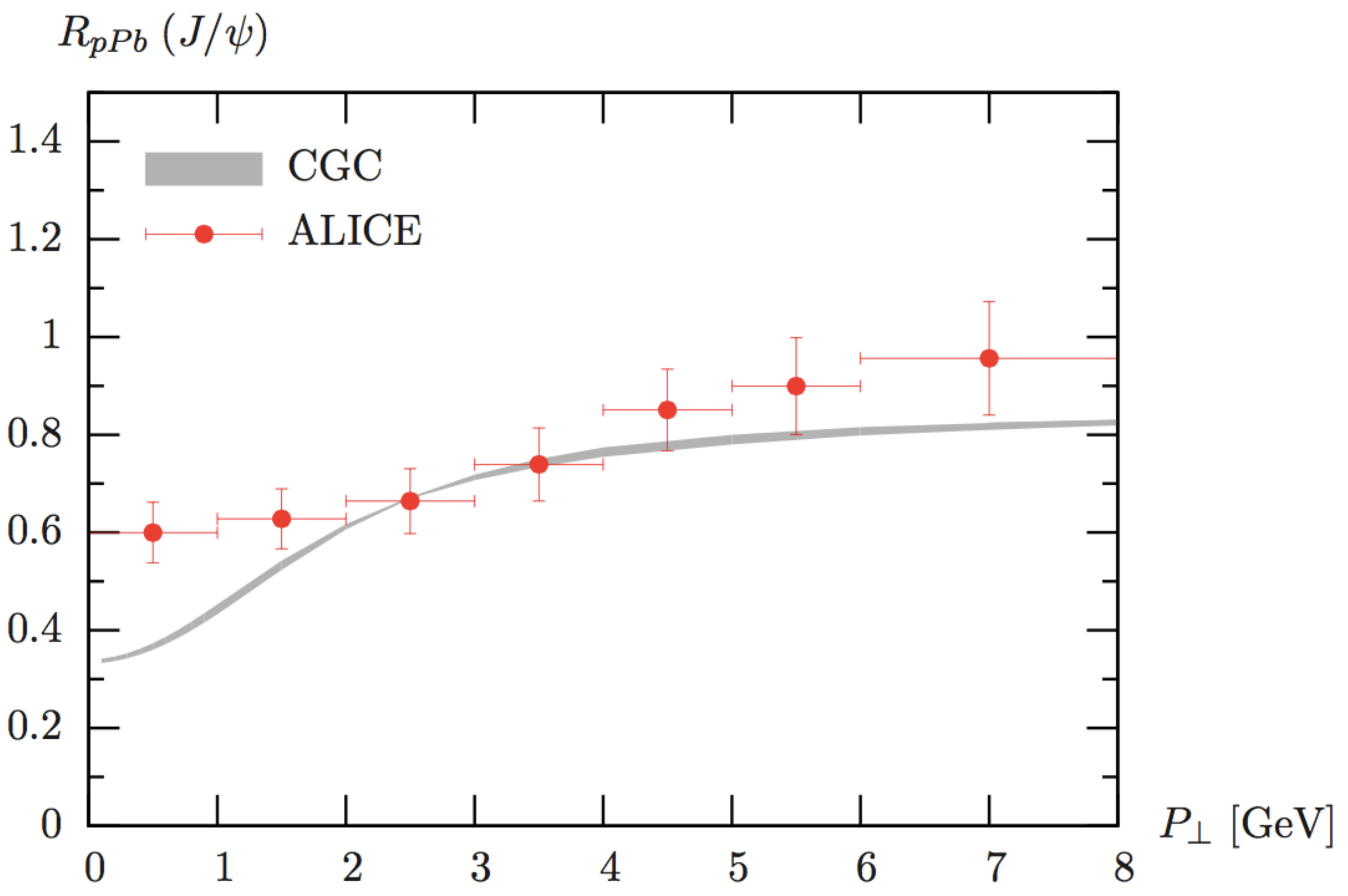}
\caption{Improved predictions of $R_{{\rm p}A}$ of $J/\psi$ production at LHC in the small-$x$ formalism with the CEM by using Eq.~(\ref{eq:rcBK-IC-Glauber}). The figures are from ~\cite{Ducloue:2015gfa}.}
\label{fig:RpA-LHC-CEM-inproved-Glauber}
\end{figure}

\begin{figure}
\centering
  \includegraphics[width=7.5cm]{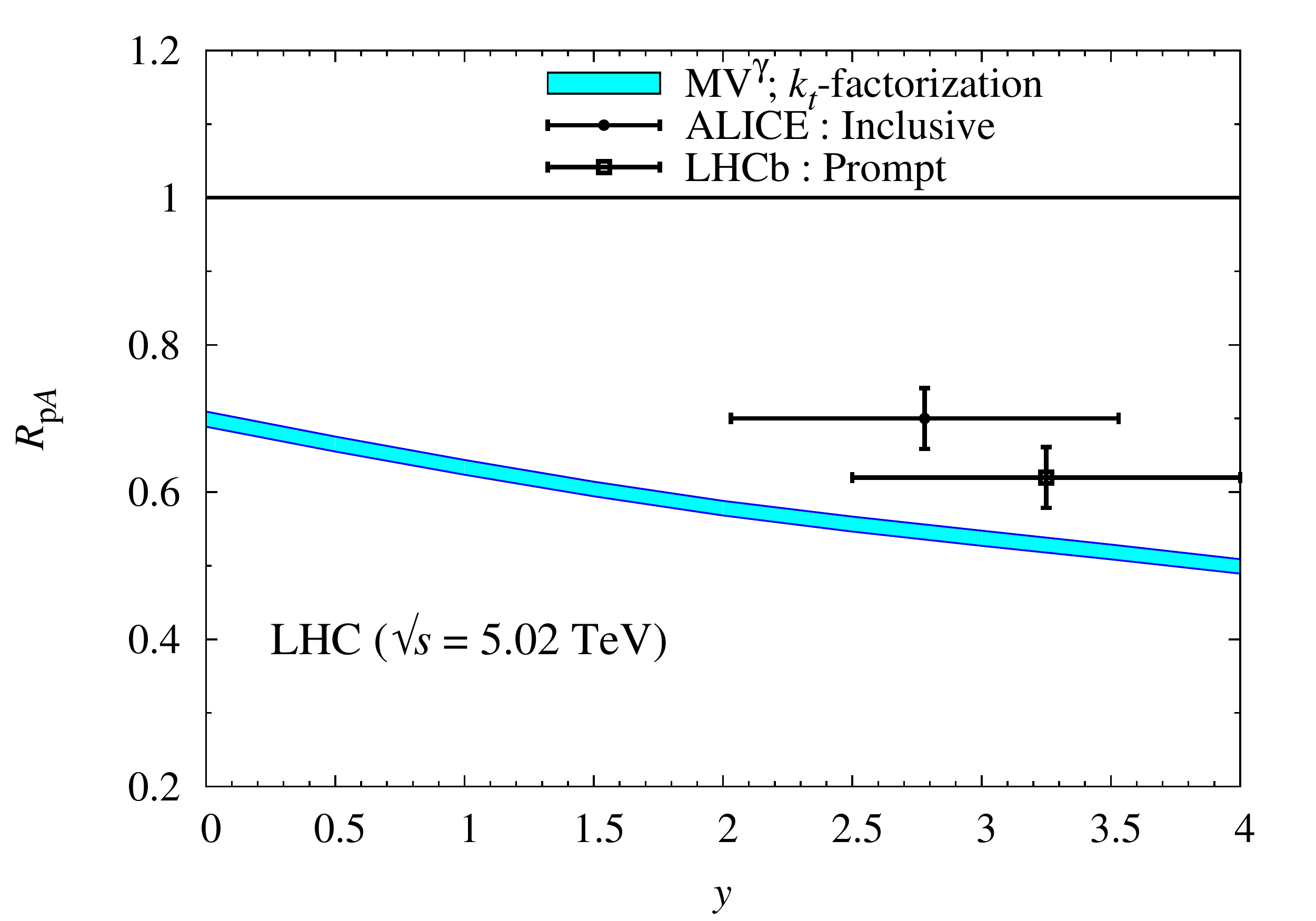}
  \includegraphics[width=7.5cm]{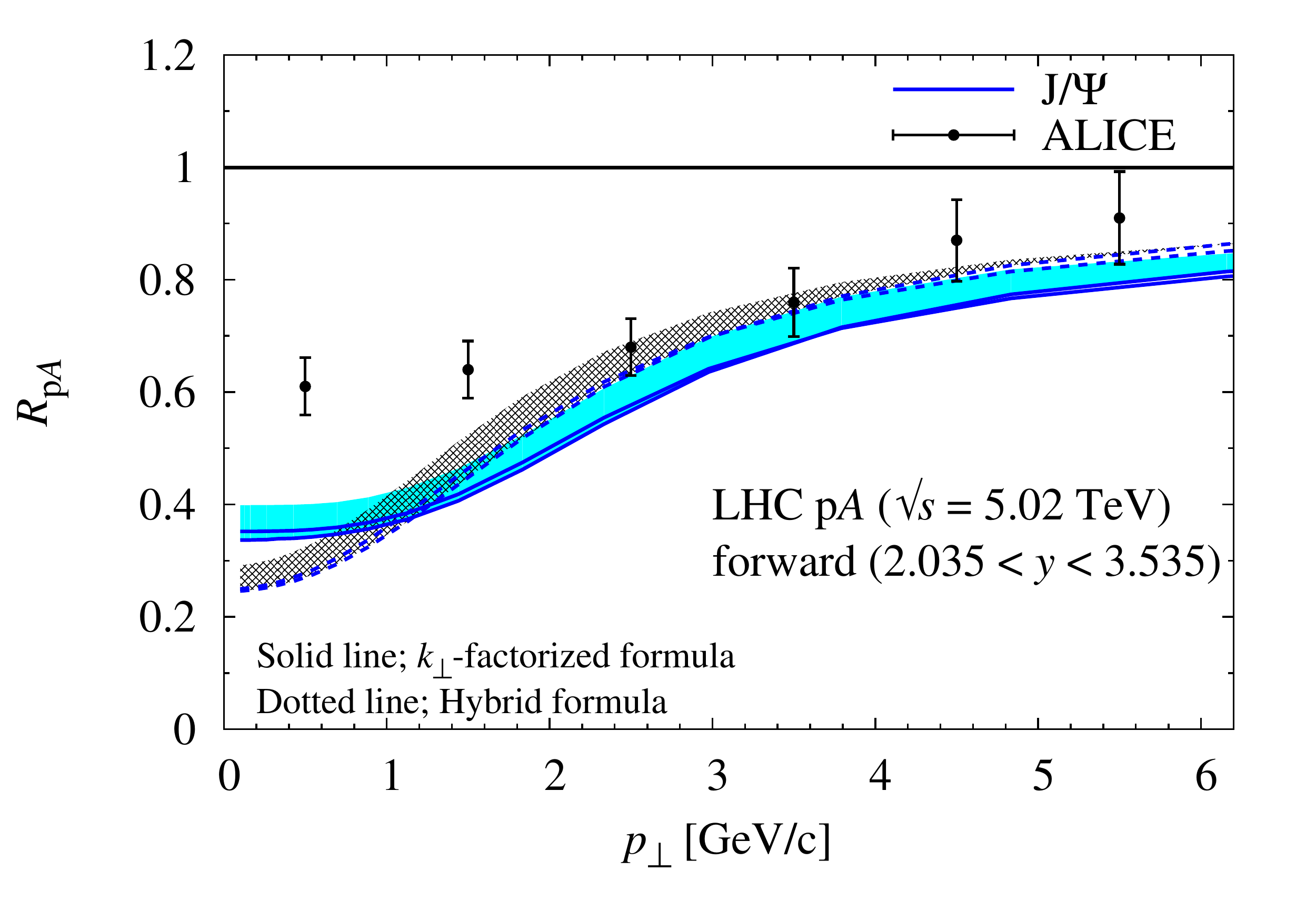}
\caption{Improved predictions of $R_{{\rm p}A}$ of $J/\psi$ production at LHC in the small-$x$ formalism incorporated with the CEM by setting $c=0.5$ in Eq.~(\ref{eq:rcBK-IC-nucleus})~\cite{Fujii:2015lld}.}
\label{fig:RpA-LHC-CEM-inproved}
\end{figure}

\begin{figure}
\centering
  \includegraphics[width=7.5cm]{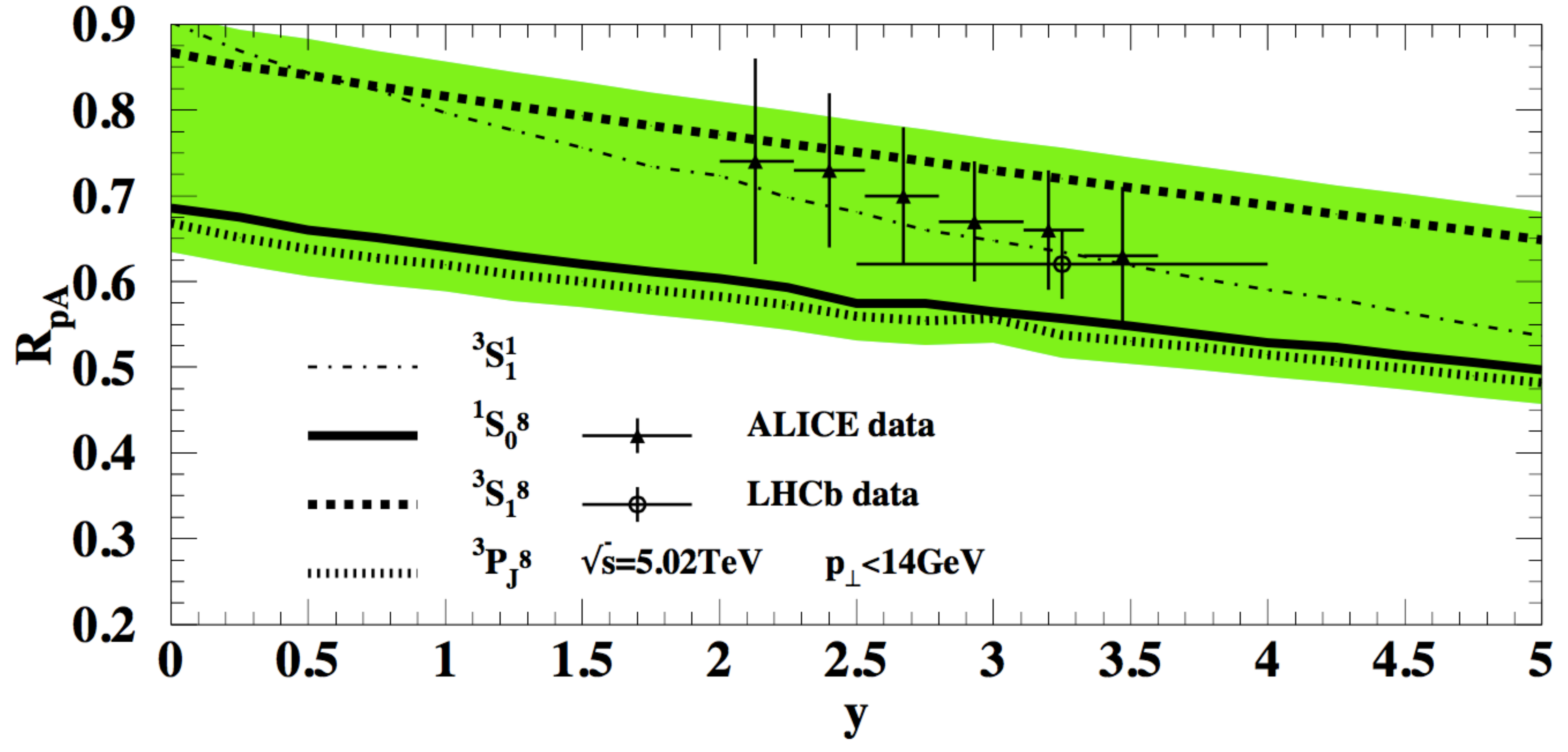}
  \includegraphics[width=7.5cm]{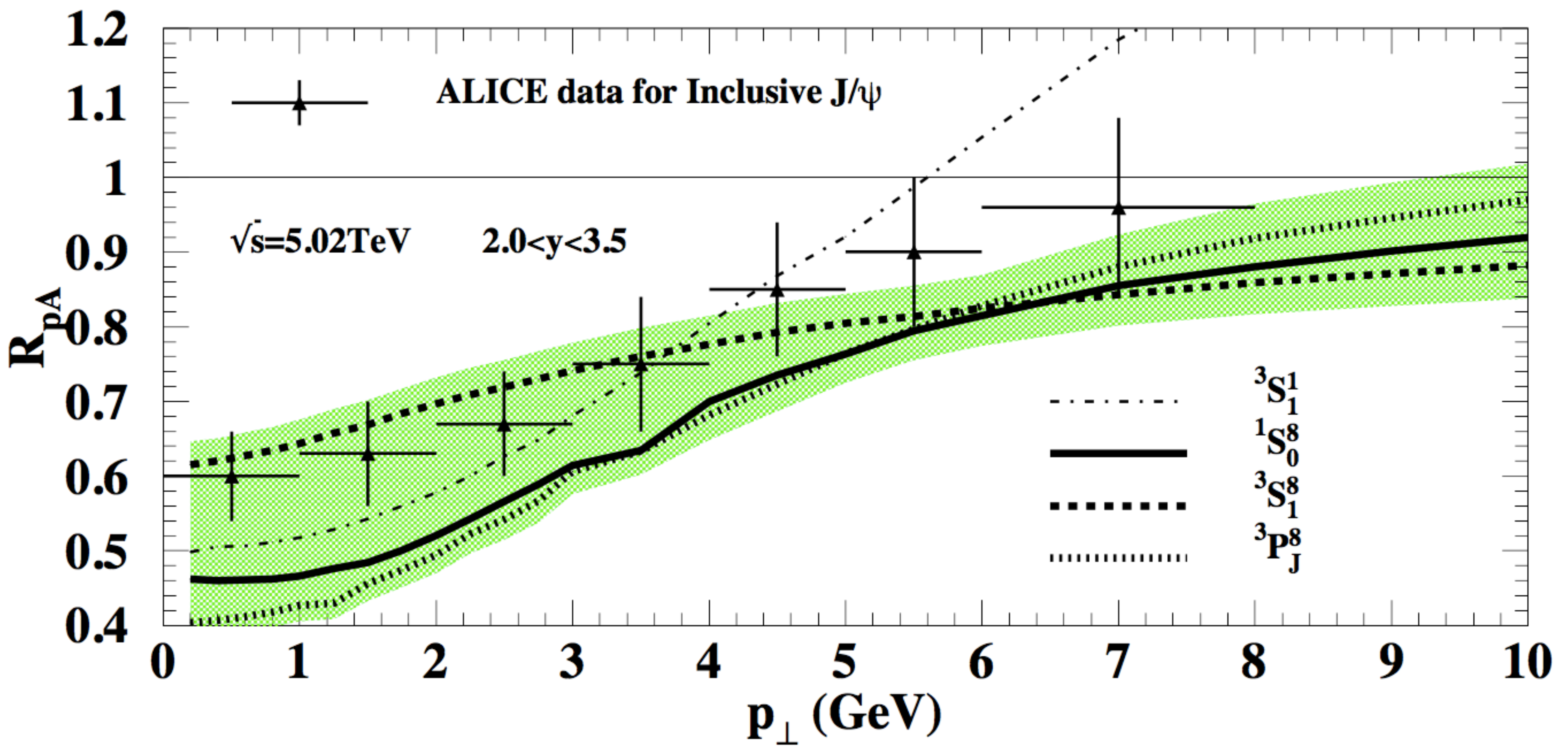}
\caption{$R_{{\rm p}A}$ of $J/\psi$ production in the small-$x$ formalism incorporated with the NRQCD LDMEs. The figures are taken from Ref.~\cite{Ma:2015sia}. }
\label{fig:NRQCD-Pt-RpA}
\end{figure}

\quad
The small-$x$ formalism contains some uncertainties in association with the input parameters, the initial condition for the rcBK equation, and the LDMEs. These uncertainties can be reduced in nuclear modification factor:
\begin{align}
R_{{\rm p}A}=\frac{1}{A}\frac{d^3\sigma_{{\rm p}A}/d^2P_\perp dy}{d^3\sigma_{\rm pp}/d^2P_\perp dy}.
\end{align}
We show in Fig.~\ref{fig:RpA-LHC-prediction} the genuine predictions for the $R_{{\rm p}A}$ of $J/\psi$ production in p$A$ collisions at LHC in the small-$x$ formalism with the CEM~\cite{Fujii:2013gxa}. In Ref.~\cite{Fujii:2013gxa}, Eq.~(\ref{eq:rcBK-IC-nucleus}) with $c=1$ is used for the rcBK initial condition for the target nucleus\footnote{$Q_{s0,A}^2=(4-6)Q_{s0}^2$ is used for numerical calculations in Ref.~\cite{Fujii:2013gxa}.}. Surprisingly, the naive evaluation in the small-$x$ formalism incorporated with the CEM provides a strong suppression of the $R_{pA}$ for forward $J/\psi$ production at LHC. 

In order to understand the strong $J/\psi$ suppression in p$A$ collisions in the small-$x$ formalism with the CEM, the impact parameter dependence, which is missed in the early results in the small-$x$ formalism, should be studied. Originally, the impact parameter dependence has been assumed to be weak for simplicity in the small-$x$ formalism. However, in fact, the gluon density in the transverse plane of the nucleus must be small at the edge of the nucleus. Therefore, if we take into account the gluon density in the nucleus correctly, the saturation effect can be weak or moderate than the naive estimate by Eq.~(\ref{eq:rcBK-IC-nucleus}) with $c=1$. 

Fig.~\ref{fig:RpA-LHC-CEM-inproved-Glauber} shows the numerical results in the Hybrid formula incorporated with the CEM by using the optical Glauber model which allows us to take into account the impact parameter dependence explicitly~\cite{Ducloue:2015gfa}. The improved results of the $R_{{\rm p}A}$ for forward $J/\psi$ production are more close to the LHC data compared to the early predictions. As explained in Section~\ref{sec:4}, they effectively set the $N_{coll}$, which is embedded in the initial condition for the rcBK equation for the nucleus, to be small for the minimum bias event due to the small value of $\sigma_0/2$. Interestingly, by using Eq.~(\ref{eq:rcBK-IC-nucleus}) with $c=0.5$ for the initial condition for the rcBK equation\footnote{$Q_{s0,A}^2=3Q_{s0}^2$ is used in Ref.~\cite{Fujii:2015lld}.}, the similar nuclear suppressions are obtained as shown in Fig.~\ref{fig:RpA-LHC-CEM-inproved}.

In addition to the results obtained in the small-$x$ formalism incorporated with the CEM, it is interesting to compare the results in the small-$x$ formalism with the NRQCD LDMEs with the LHC data. As checked in Refs.~\cite{Ma:2014mri,Ma:2015sia}, the color singlet channel does not contribute to the total cross section for $J/\psi$ production in pp and p$A$ collisions. Therefore, we can expect that the nuclear suppressions of the color octet contributions in the small-$x$ formalism with the NRQCD factorization is the same as the one in the small-$x$ formalism with the CEM.

Fig.~\ref{fig:NRQCD-Pt-RpA} shows the $R_{{\rm p}A}$ for forward $J/\psi$ production at LHC computed in the small-$x$ formalism incorporated with the NRQCD LDMEs~\cite{Ma:2015sia}. Although it is hard to determine the most dominant channel for $J/\psi$ production, one finds that the individual channels are in good agreement with the LHC data within the large uncertainties in association with the NRQCD LDMEs. The $P_\perp$ distribution of the $R_{pA}$ for $J/\psi$ production depends on the quantum state of the produced $q\bar q$ pair in the intermediate state, since each of the channels has its own hard matrix element. In fact, the moderate $J/\psi$ suppression in the small-$x$ formalism with the NRQCD LDMEs comes from the fact that the MV initial condition for the rcBK equation with $Q_{s0,A}^2=2Q_{s0}^2$ is used for the target nucleus~\cite{Ma:2015sia}.

From all the above results obtained in the small-$x$ formalism incorporated with the CEM and also the NRQCD LDMEs, we can conclude that we need to at least set the small value of the saturation scale for the target nucleus to describe the LHC data of forward $J/\psi$ production in p$A$ collisions.

\subsection{Event activity dependence}

\begin{figure}
\centering
  \includegraphics[width=7.5cm]{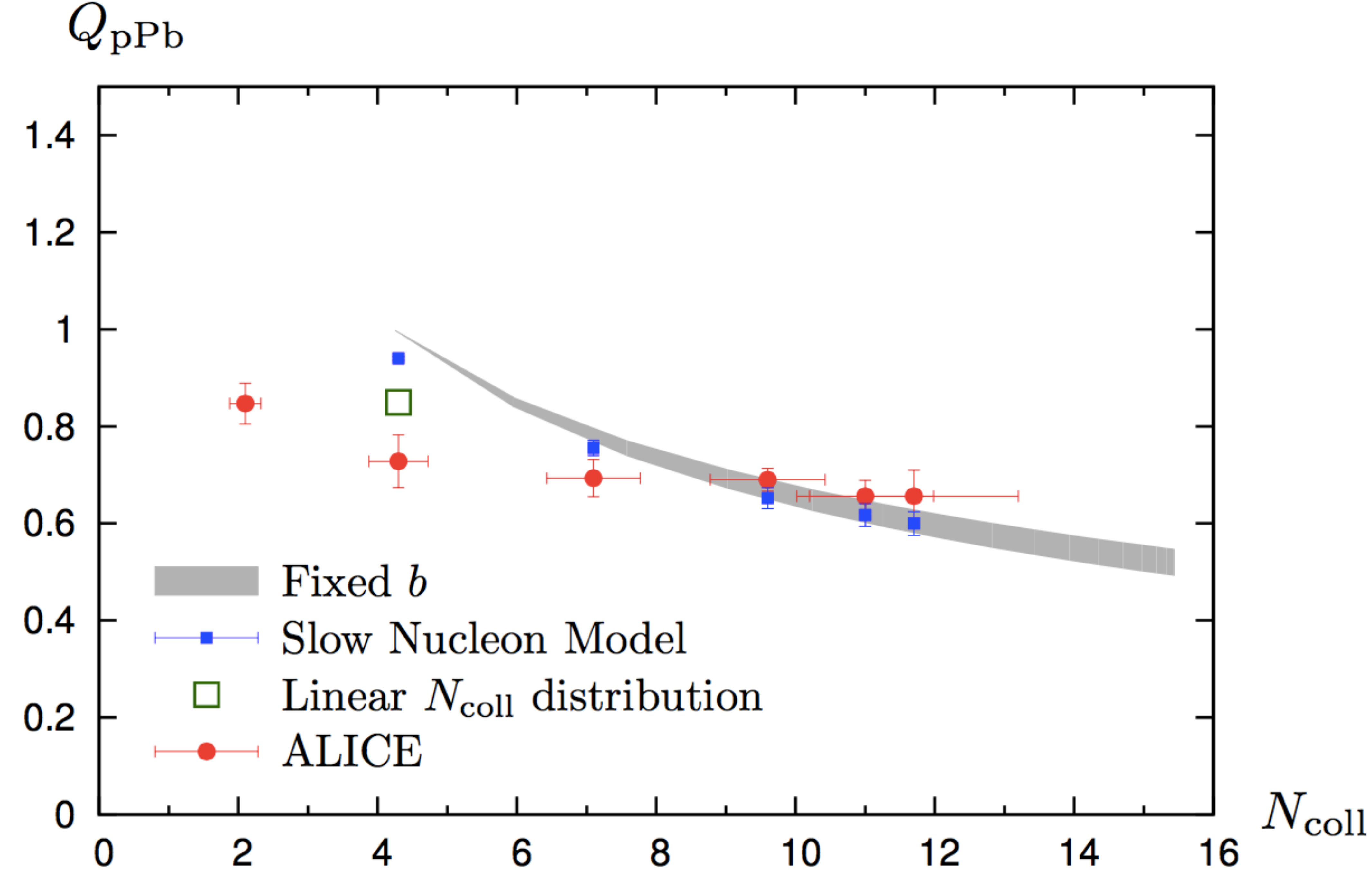}
  \includegraphics[width=7.5cm]{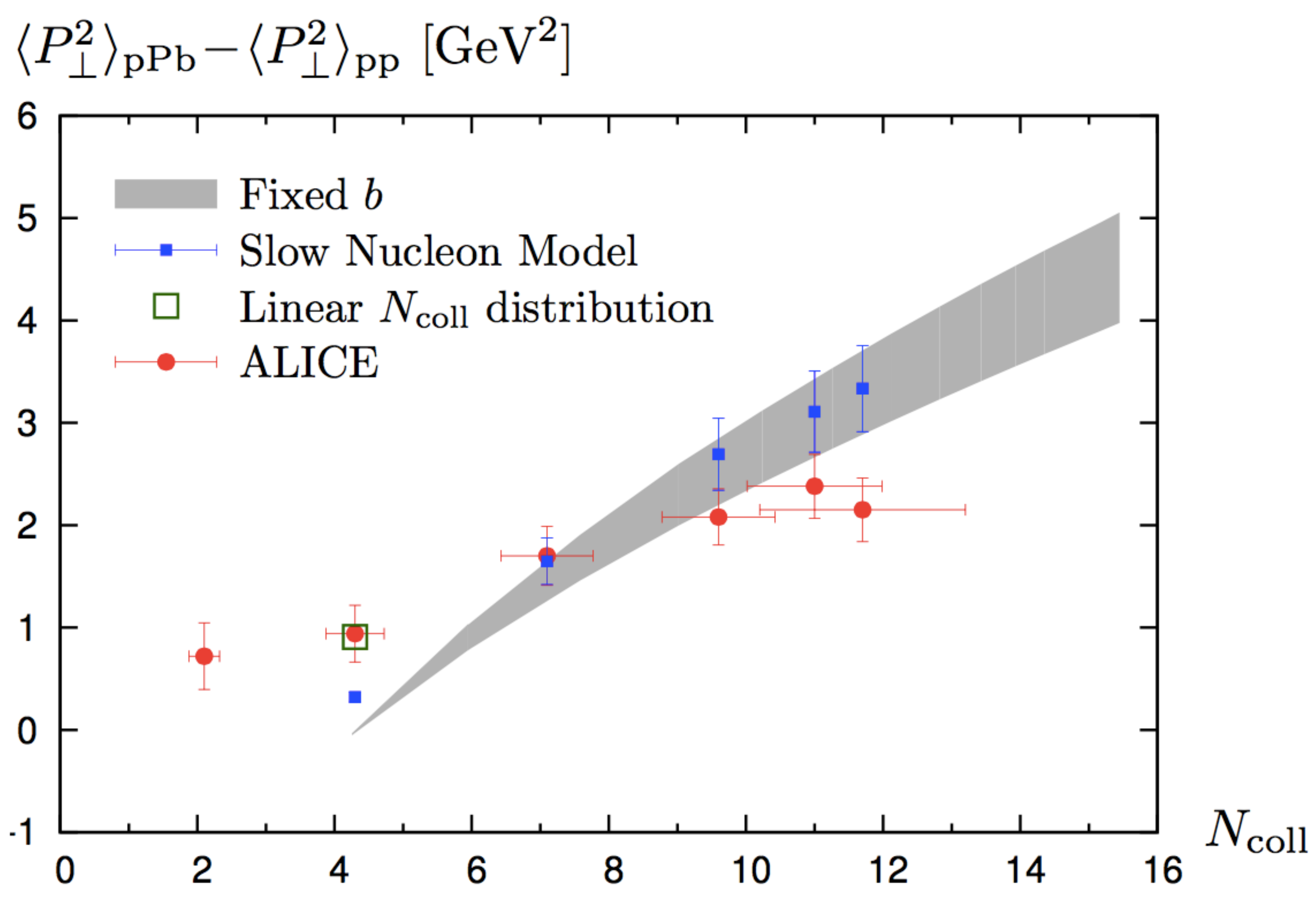}
\caption{Event activity dependence of $Q_{{\rm p}A}$ and $\Delta\langle P_\perp^2\rangle_{{\rm p}A}$ for forward $J/\psi$ production in p$A$ collisions at LHC. Figures are taken from Ref.~\cite{Ducloue:2016pqr}.}
\label{fig:Ncoll-dep}
\end{figure}

\quad
In addition to minimum bias events, event activity dependence (or centrality dependence) of $J/\psi$ production in p$A$ collisions is also useful to examine the gluon saturation dynamics in the target nucleus. Particularly,  correct understanding of the impact parameter dependence of the saturation effect in p$A$ collisions enables us to study the underlying physics behind the moderate $J/\psi$ suppression in p$A$ collisions at LHC at forward rapidity.

The nuclear modification factor for each event activity in p$A$ collision is defined as
\begin{align}
Q_{{\rm p}A}=\frac{1}{A\langle T_A\rangle }\frac{d^3N_{{\rm p}A}/d^2P_\perp dy}{d^3\sigma_{\rm pp}/d^2P_\perp dy}
\end{align}
where $dN_{{\rm p}A}$ is the invariant yield for $J/\psi$ production in p$A$ collisions and $\langle T_A\rangle$ is related to the number of binary collision $N_{\rm coll}$, namely, the centrality class~\cite{Adam:2015jsa}. 
It is naively anticipated that the nuclear suppression of forward $J/\psi$ production in p$A$ collisions should be more pronounced at large $N_{\rm coll}$ rather than small $N_{\rm coll}$. In the optical Glauber model, large $N_{\rm coll}$ corresponds to central p$A$ collision while small $N_{\rm coll}$ corresponds to peripheral collision, although Monte Carlo Glauber model does not provide such a simple picture.

Fig.~\ref{fig:Ncoll-dep} (left) shows the $Q_{{\rm p}A}$ of forward $J/\psi$ production at LHC as a function of $N_{\rm coll}$ computed in the small-$x$ formalism incorporated with the CEM and Eq.~(\ref{eq:rcBK-IC-Glauber})~\cite{Ducloue:2016pqr}. The theoretical results strongly depend on the value of $N_{\rm coll}$ in p$A$ collision. This is because the Wood-Saxon nucleon distribution used in the optical Glauber model gives the strong impact parameter dependence of $N_{\rm coll}$. However, the LHC data shows the $N_{\rm coll}$ distribution of the $Q_{pA}$ is rather gradual than the theoretical results. 

Fig.~\ref{fig:Ncoll-dep} (right) shows the $N_{\rm coll}$ dependence of the $P_\perp$ broadening of forward $J/\psi$ production at LHC. One finds that the deviation $\Delta\langle P_\perp^2\rangle_{{\rm p}A}=\langle P_\perp^2\rangle_{{\rm p}A}-\langle P_\perp^2\rangle_{{\rm pp}}$ strongly depends on the value of the $N_{\rm coll}$ while the LHC data shows the $N_{\rm coll}$ dependence of the $\Delta\langle P_\perp^2\rangle_{{\rm p}A}$ is not so large in fact.

We must address this issue in the future by improving the small-$x$ formalism. Of particular important point is that we must reproduce both the smaller saturation scale of the nucleus in the minimum bias event and the gradual change of the $N_{\rm coll}$ as the centrality changes.


\section{Sudakov implementation in the Small-$x$ formalism}\label{sec:7}

\quad
In principle, $\Upsilon$ production in p$A$ collisions can be calculated in the small-$x$ formalism by incorporating with the CEM or the NRQCD LDMEs, since the discussion in Section~\ref{sec:2} holds true for $\Upsilon$. One expects that $P_\perp$ distribution of $\Upsilon$ production in pp or p$A$ collisions should be similar to the one for $J/\psi$ due to the saturation effects with similar kinematics and physical input parameters between $\Upsilon$ and $J/\psi$, although the mass scale of $\Upsilon$ is different from the one of $J/\psi$. However, the small-$x$ formalism cannot describe the $P_\perp$ distribution of $\Upsilon$ production in pp at LHC even in the very forward rapidity region~\cite{Fujii:2013gxa,Ducloue:2015gfa}. Indeed, the LHCb data~\cite{LHCb:2012aa} shows that the mean $P_\perp$ of $\Upsilon$ in pp collisions is much larger than that of $J/\psi$~\cite{Aaij:2011jh} for a wide range of rapidity windows.

This fact does not implies that the small-$x$ formalism is completely wrong or cannot be applied to $\Upsilon$ production. Here, we must realize that the large mass scale of $\Upsilon$ should open a large phase space for $\Upsilon$ production at $M\gg P_\perp$. The large phase space can allow us to take into account parton shower (soft gluons emission) effect in initial state. For $\Upsilon$ production, parton shower effect can be expressed as Sudakov double logarithmic corrections, $\frac{\alpha_sN_c}{2\pi}\ln^2\frac{M^2}{P_\perp^2}$~\cite{Qiu:2013qka,Berger:2004cc,Sun:2012vc}. The Sudakov double logarithmic corrections can be resummed by using Colins-Soper-Sterman (CSS) formalism~\cite{Collins:1984kg}, while the small-$x$ logarithmic correction is $\frac{\alpha_sN_c}{2\pi^2}\ln\frac{1}{x}$ which can be resummed using the BK equation. Thus, we must take into account these large corrections simultaneously for $\Upsilon$ production at low $P_\perp$ in the forward rapidity.

Now we go through $\Upsilon$ production in pp collision in the small-$x$ formalism incorporated with the CEM and the Sudakov factor. In practice, by taking into account the parton shower effect in the small-$x$ formalism, the initial gluon distribution for the projectile proton can be modified. Therefore, the differential cross section for forward $q\bar q$ pair production in pp collisions in the Hybrid formula can be cast into~\cite{Watanabe:2015yca}
\begin{align}
\frac{d\sigma_{q\bar{q}}}{d^2q_{\perp} d^2p_{\perp} dy_q dy_{p}}
=
\frac{\alpha_sN_cS_{p\perp} }{64\pi^2 C_F}
\int d^2l_\perp d^2k_\perp
\Xi_{\rm coll}(k_{2\perp},k_{\perp}-zl_\perp)
F_{\rm TMD}(l_\perp)
F_{Y_{\rm p^\prime}}\left(k_\perp\right)
F_{Y_{\rm p^\prime}}\left(k_{2\perp}-k_\perp+l_\perp\right)
\label{eq:xsection-sudakov}
\end{align}
where we have already assumed the large-$N_c$ approximation. $\Xi_{\rm coll}$ is the hard scattering part in the Hybrid formula. In the above expressions, we have introduced the momentum fraction $z$ for the quark $z={q^+}/({q^++p^+})$ and $z_{p}=1-z$ for the antiquark. The new transverse momentum dependent (TMD) gluon distribution function for the projectile proton is given by
\begin{align}
F_{\rm TMD}(M,l_\perp)=\int \frac{d^2b_\perp}{(2\pi)^2}e^{-ib_\perp\cdot l_\perp} e^{-S_{\rm Sud}(M,b_\perp)}x_{\rm p}G\left(x_{\rm p},\mu=\frac{c_0}{b_\perp}\right)
\end{align}
where $x_{\rm p}G$ is the collinear gluon distribution function for the projectile proton at momentum fraction $x_{\rm p}$. 

In the CSS formalism, the Sudakov factor is separated from the perturbative part and the nonperturbative part as $S_{\rm Sud}(M,b)=S_{\rm perp}(M,b_\star)+S_{\rm NP}(M,b)$ by using $b_\star$-prescription: $b_\star=b/\sqrt{1+(b/b_{\rm max})^2}$. 
The perturbative part of the Sudakov factor for $b_\star\sim b\ll b_{\rm max}$ is given by
\begin{align}
S_{\rm perp}(M,b)=\int_{c_0/b^2}^{M^2}\frac{d\mu^2}{\mu^2}\left[A\ln\left(\frac{M^2}{\mu^2}\right)+B\right]
\end{align}
where the coefficient functions $A$ and $B$ can be calculated perturbatively as $A=\sum\limits_{i=1} A^{(i)}\left(\frac{\alpha_s}{\pi}\right)^i$ and 
$B=\sum\limits_{i=1} B^{(i)}\left(\frac{\alpha_s}{\pi}\right)^i$.
For the one loop correction, $A^{(1)}=C_A$ and $B^{(1)}=-(b_0+\frac{1}{2}\delta_{8c})N_c$
where $b_0=\left(\frac{11}{6}N_c-\frac{n_f}{3}\right)\frac{1}{N_c}$.
As for $B^{(1)}$, the factor $\delta_{8c}$ is significant only in the production of a color octet $q\bar q$. The nonperturbative Sudakov factor for $b>b_{\rm max}$ can be determined by data fitting. In Ref.~\cite{Sun:2012vc}, the nonperturbative Sudakov factor is determined with the following functional form:
\begin{align}
S_{\rm NP}(M,b)=\exp\left[\frac{b^2}{2}\left(-g_1-g_2\ln\left(\frac{M}{2Q_0}\right)-g_1g_3\ln(100x_{\rm p}x_{\rm p^\prime})\right)\right]
\label{eq:npsudakov}
\end{align}
where the parameters $g_1$, $g_2$, $g_3$, $Q_0$, and $b_{\rm max}$ are found in Ref.~\cite{Sun:2012vc}.

\begin{figure}
\centering
  \includegraphics[width=7.5cm]{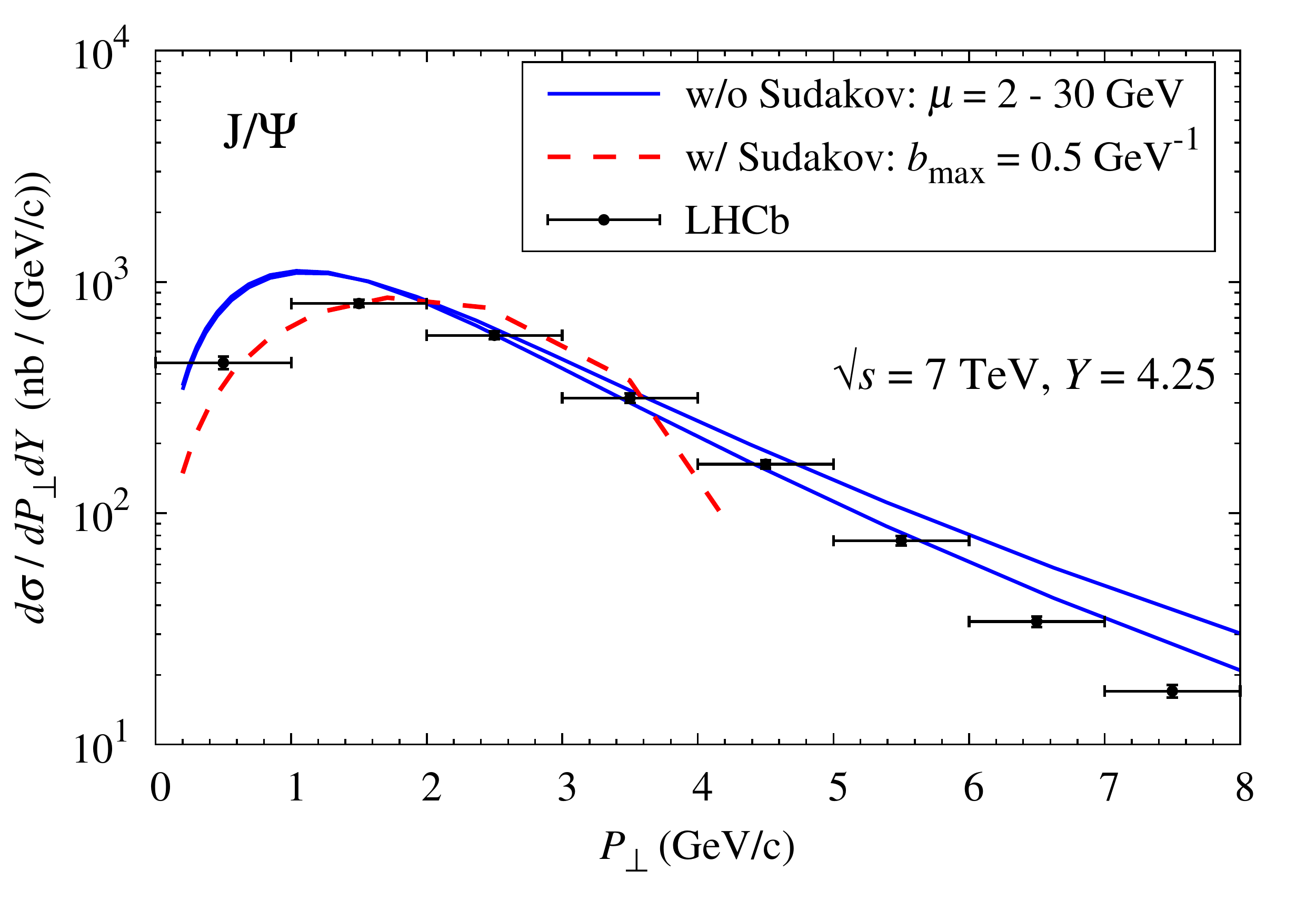}
  \includegraphics[width=7.5cm]{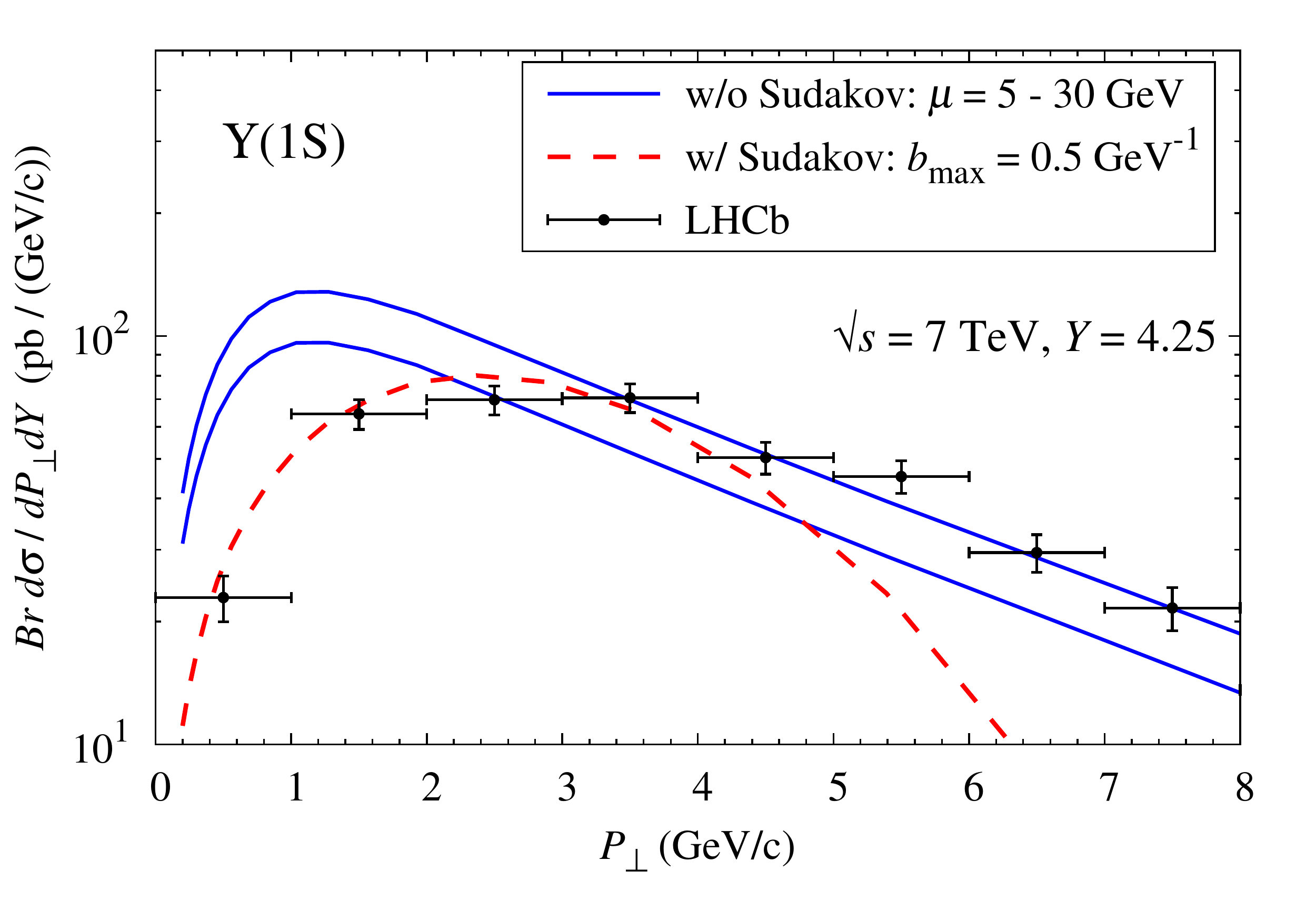}
\caption{$P_\perp$ distributions of $J/\psi$ and $\Upsilon$ production in pp collisions at $\sqrt s=7$\;TeV at $Y=4.25$. Blue solid (Red dashed) lines are the results in the small-$x$ formalism without (with) the Sudakov factor.}
\label{fig:sudakov}
\end{figure}

Fig.~\ref{fig:sudakov} shows the numerical results for $J/\psi$ and $\Upsilon$ production in pp collisions at $\sqrt s=7$\;TeV at forward rapidity in the small-$x$ formalism incorporated with the CEM~\cite{Watanabe:2015yca}. As obtained in Refs.~\cite{Fujii:2013gxa,Ducloue:2015gfa}, the forward $\Upsilon$ production cannot be described in the small-$x$ formalism without the Sudakov factor. On the other hand, the calculation with the parton shower effect using Eq.~(\ref{eq:xsection-sudakov}) reproduces the LHC data points of $\Upsilon$ production in pp collisions. The parton shower effect also modifies the $P_\perp$ distribution of $J/\psi$ production in pp collisions slightly.

Of particular importance is that the parton shower effect is dominant than the saturation effect for $\Upsilon$ production in pp collisions but less pronounced for $\Upsilon$ production in p$A$ at forward rapidity, since the saturation scale for the target nucleus is larger than the one for the proton. Clearly speaking, although $\Upsilon$ production is still an interesting probe to investigate the gluon saturation dynamics in hadron and nucleus, the parton shower effect is indispensable in order to consistently describe $\Upsilon$ productions in the small-$x$ formalism. The same can be said of $J/\psi$ production.

We comment on the uncertainty bands of the numerical results shown in Fig.~\ref{fig:sudakov}. The factorization scale $\mu$ is chosen between 2 or 5 GeV to 30 GeV in the small-$x$ formalism without the Sudakov factor. On the other hand, by incorporating the CSS formalism with the small-$x$ formalism, one can choose $\mu=c_0/b_\perp$ in the coordinate space. Therefore, no uncertainty in association with the choice of $\mu$ is shown in Fig.~\ref{fig:sudakov} for the results obtained in the small-$x$ formalism with the Sudakov factor.

In addition, one must keep in mind that so-called $Y$-term is missed in this calculation~\cite{Boglione:2014oea}. Necessarily, $F_{\rm TMD}$ becomes negative at $P_\perp\gtrsim M$ in which the CSS formalism is no longer valid. Indeed, one finds in Fig.~\ref{fig:sudakov} that the $P_\perp$ spectrums of $J/\psi$ and $\Upsilon$ with the Sudakov factor decreases rapidly at large $P_\perp$ compared to the fixed order results without the Sudakov factor. The $Y$-term is required to connect the small-$x$ formalism incorporated with the Sudakov factor to the fixed order calculation which is responsible for the large $P_\perp$ region, although we should switch the small-$x$ formalism to the collinear factorization framework at large $P_\perp$.

\section{Summary}\label{sec:8}

\quad
In this paper, we have gone through $J/\psi$ and $\Upsilon$ production in pp and p$A$ collisions at RHIC and LHC in the small-$x$ formalism. The small-$x$ formalism is originally derived for a dilute-dense system of order ${\cal O}(\rho^1_p\rho_A^\infty)$ which accords with p$A$ collision and pp collision at forward rapidity.
The concept of the effective factorization between the short distance hard scattering part and the long distance bound state formation process allows us to apply the small-$x$ formalism to the $q\bar q$ pair production in p$A$ collisions and pp collisions, although a systematic description of bound state formation depends on the model at present.

As shown in Section~\ref{sec:6}, the small-$x$ formalism can describe  the $P_\perp$ spectrum for forward $J/\psi$ production at $P_\perp\lesssim Q_s$ in pp and p$A$ collisions at RHIC and LHC by incorporating either the CEM or the NRQCD LDMEs, although the small-$x$ formalism cannot predict the normalization due to the large uncertainties, including the input parameters, the initial condition for the rcBK equation. The consequence of the small-$x$ formalism with the NRQCD approach is that the color octet channels dominate in $J/\psi$ production at low $P_\perp$ both in pp and p$A$ collisions.

The nuclear modification factor of $J/\psi$ for minimum bias event and several centrality classes in p$A$ collision almost do not depend on the input parameters except for the initial condition for the rcBK equation for the target nucleus. The previous papers~\cite{Ducloue:2015gfa,Ma:2015sia,Fujii:2015lld} suggested that the initial saturation scale for the target nucleus for minimum bias event should be small value ($Q_{sA}^2=(2-3)Q_{sp}^2$) to reproduce the $R_{pA}$ for $J/\psi$ production at LHC in the forward rapidity region. In the mean time, we need to model further the initial condition for the rcBK equation for the target nucleus to describe the $N_{coll}$ dependence of the $Q_{pA}$. These improvements certainly require an approach beyond the optical Glauber model.

For $\Upsilon$ production, the parton shower effect is indispensable to interpret data. For $J/\psi$ production, the parton shower effect could be also important, however, we may have to consider nonperturbative effect at low $P_\perp$ for $J/\psi$ production. Thus far, the parton shower effect on quarkonium production has been studied in the small-$x$ formalism incorporated with the CEM. It is interesting to perform the same calculation in the small-$x$ formalism with the NRQCD LDMEs. In order to address this calculation, we should also conduct full NLO calculations for the $q\bar q$ pair production in p$A$ collisions in the small-$x$ formalism. 

Finally, it is worth giving comments on $\psi(2S)$ production in the small-$x$ formalism. Interestingly, for $\psi(2S)$ production, final state interaction can play an important role to describe data, since $\psi(2S)$ is the loosely bound state. In fact, the LHC data shows that the nuclear modification factor for $\psi(2S)$ production in p$A$ collisions at forward rapidity is strongly suppressed than that of $J/\psi$ production~\cite{Abelev:2014zpa}. The saturation effect must be more or less the same between $J/\psi$ and $\psi(2S)$. Therefore, the data indicates that the so-called comover interaction in final state can affect $\psi(2S)$ production~\cite{Ferreiro:2014bia}. The final state interaction is likely to violate the factorization between the hard part at short distance and the bound state formation part at long distance. Nevertheless, by taking into account both the saturation effect and the comover effect simultaneously, the nuclear suppressions of $J/\psi$ and $\psi(2S)$ can be described well in the small-$x$ formalism. The details of a calculation of $\psi(2S)$ production in the small-$x$ formalism will be reported separately~\cite{psi2s-cgc}.

\begin{acknowledgements}
This work is supported by the National Science Foundation of China (NSFC) under Grant \#11575070 and Jefferson Science Associates,
LLC under  U.S. DOE Contract \#DE-AC05-06OR23177
and U.S. DOE Grant \#DE-FG02-97ER41028. 
I would like to thank F.~Arleo, B.~Duclou$\acute{\text{e}}$, J.-P.~Lansberg, T.~Lappi, H.~Fujii, Y.-Q.~Ma, J.-W.~Qiu,  R.~Venugopalan, B.-W.~Xiao, F.~Yuan for interesting discussions and comments. 

The final publication is available at Springer via http://dx.doi.org/10.1007/s00601-017-1297-z.
\end{acknowledgements}



\end{document}